\documentclass[%
 reprint,
nofootinbib,
 amsmath,amssymb,
 aps,
prb,
]{revtex4-2}
\let\revappendix\appendix

\usepackage{graphicx}
\usepackage{dcolumn}
\usepackage{bm}
\usepackage{hyperref}
\usepackage{mathtools}
\usepackage{empheq}
\usepackage{esint}
\usepackage[dvipsnames,table]{xcolor}
\usepackage{makecell,tabularx,booktabs}
\usepackage{siunitx}

\setcellgapes{2.5pt}
\newcolumntype{Y}{>{\centering\arraybackslash}X}

\usepackage[ruled,vlined,linesnumbered]{algorithm2e}

\SetCommentSty{mycommfont}

\usepackage{cleveref}
\usepackage{enumitem}
\setitemize{leftmargin=*,noitemsep,topsep=2pt,parsep=1.8pt,partopsep=2pt}
\setenumerate{leftmargin=*,noitemsep,topsep=2pt,parsep=2pt,partopsep=2pt}

\newcommand{\parder}[2]{\frac{\partial #1}{\partial #2}}

\newcommand{\del}{\nabla}
\newcommand{\bPsi}{\boldsymbol{\Psi}}

\newcommand{\bvPsi}{\boldsymbol{\varPsi}}

\newcommand{\ncsocfe}{\texttt{NCSOC-DFT-FE}}
\newcommand{\qe}{\texttt{QE}}

\newcommand{\bg}{\boldsymbol{g}}

\newcommand{\bk}{\boldsymbol{k}}

\newcommand{\bmm}{\boldsymbol{m}}

\newcommand{\br}{\boldsymbol{{r}}}

\newcommand{\bu}{\boldsymbol{u}}

\newcommand{\bx}{\boldsymbol{\textbf{x}}}
\newcommand{\by}{\boldsymbol{\textbf{y}}}
\newcommand{\bz}{\boldsymbol{\textbf{z}}}

\newcommand{\bB}{\boldsymbol{B}}

\newcommand{\bD}{\boldsymbol{D}}
\newcommand{\bE}{\boldsymbol{\textbf{E}}}

\newcommand{\bRcal}{\boldsymbol{\mathcal{R}}}
\newcommand{\bH}{\boldsymbol{\textbf{H}}}
\newcommand{\btH}{\widetilde{\bH}}
\newcommand{\bPt}{\boldsymbol{\textbf{P}}}
\newcommand{\bI}{\boldsymbol{I}}
\newcommand{\bJ}{\boldsymbol{\textbf{J}}}
\newcommand{\bK}{\boldsymbol{\textbf{K}}}

\newcommand{\bX}{\boldsymbol{\textbf{X}}}
\newcommand{\bY}{\boldsymbol{\textbf{Y}}}
\newcommand{\bL}{\boldsymbol{\textbf{L}}}

\newcommand{\bM}{\boldsymbol{\textbf{M}}}
\newcommand{\bN}{\boldsymbol{\textbf{N}}}

\newcommand{\bQ}{\boldsymbol{\textbf{Q}}}
\newcommand{\bR}{\boldsymbol{\textbf{R}}}

\newcommand{\bU}{\boldsymbol{\textbf{U}}}
\newcommand{\bV}{\boldsymbol{{\textbf{V}}}}

\newcommand{\bLam}{\boldsymbol{\Lambda}}
\newcommand{\bDel}{\boldsymbol{\Delta}}

\newcommand{\order}{\mathcal{O}}

\newcommand{\dr}{\,d\br}
\newcommand{\dk}{\,d\bk}
\newcommand{\detb}[1]{\,\det\left(#1\right)}

\newcommand{\derveps}[1]{\frac{d #1}{d\varepsilon}}
\newcommand{\parderveps}[1]{\frac{\partial #1}{\partial\varepsilon}}
\newcommand{\dervepsZ}[1]{\frac{d #1}{d\varepsilon}\biggl|_{\varepsilon=0}}
\newcommand{\pardervepsZ}[1]{\frac{\partial #1}{\partial\varepsilon}\biggl|_{\varepsilon=0}}

\newcommand{\pardervepsb}[1]{\frac{\partial}{\partial\varepsilon}\left\{#1\right\}}
\newcommand{\pardervepsbZ}[1]{\frac{\partial}{\partial\varepsilon}\left\{#1\right\}\biggl|_{\varepsilon=0}}

\newcommand{\ket}[1]{\left| #1 \right>} 
\newcommand{\bra}[1]{\left< #1 \right|} 

\newcommand{\brho}{\boldsymbol{\rho}}%
\newcommand{\btau}{\boldsymbol{\tau}}%
\newcommand{\bUpsilon}{\boldsymbol{\Upsilon}}%
\newcommand{\bvrho}{\boldsymbol{\varrho}}%
\newcommand{\bsigma}{\boldsymbol{\sigma}}%
\newcommand{\funcder}[2]{\frac{\delta #1}{\delta #2}}%
\newcommand{\iu}{\mathrm{i}}%
\newcommand{\norm}[1]{\left\lVert#1\right\rVert}
\DeclarePairedDelimiter\abs{\lvert}{\rvert}%

\DeclareRobustCommand{\rchi}{{\mathpalette\irchi\relax}}
\newcommand{\irchi}[2]{\raisebox{\depth}{$#1\chi$}} 

\begin{document}

\preprint{APS/123-QED}

\title{Finite-element methods for noncollinear magnetism and spin-orbit coupling in real-space pseudopotential density functional theory}

\author{Nikhil Kodali}
\author{Phani Motamarri}%
 \email{phanim@iisc.ac.in}
\affiliation{%
 Department of Computational and Data Sciences, Indian Institute of Science, Bangalore 560012
}%





\begin{abstract}
We introduce an efficient finite-element approach for large-scale real-space pseudopotential density functional theory (DFT) calculations incorporating noncollinear magnetism and spin-orbit coupling effects. The approach, implemented within the open-source DFT-FE computational framework, fills a significant gap in real-space DFT calculations using finite element basis sets, which offer several advantages over traditional DFT basis sets. In particular, we leverage the local reformulation of DFT electrostatics to derive the finite-element (FE) discretized governing equations involving two-component spinors.  To evaluate the widely used GGA exchange-correlation potentials in these governing equations under the locally collinear approximation, we devise a numerical strategy that avoids the computation of ill-defined gradients of magnetization direction near zero magnetization. Additionally, we utilize an efficient self-consistent field iteration approach based on Chebyshev filtered subspace iteration procedure exploiting the sparsity of local and non-local parts of FE discretized Hamiltonian to solve the underlying nonlinear eigenvalue problem based on a two-grid strategy. Furthermore, we propose using a generalized functional within the framework of noncollinear magnetism and spin-orbit coupling with a stationary point at the minima of the Kohn-Sham DFT energy functional to develop a unified framework for computing atomic forces and periodic unit-cell stresses. Validation studies against plane-wave implementations show excellent agreement in ground-state energetics, vertical ionization potentials, magnetic anisotropy energies, band structures, and spin textures. The proposed method achieves up to 8x–11x speedups for semi-periodic and periodic systems with $\sim$5000-7000 electrons in terms of minimum wall times compared to widely used plane-wave implementations on CPUs in addition to exhibiting significant computational advantage on GPUs for material systems with as many as 20,000 electrons. The proposed approach offers a fast, scalable, and systematically convergent framework for large-scale DFT calculations accounting for noncollinear magnetism and spin-orbit coupling, enabling more complex material simulations and extending the range of \emph{ab initio} studies.
\end{abstract}


\maketitle


\section{\label{sec:intro} Introduction}

Spin-orbit coupling (SOC) and noncollinear magnetism are integral for predicting complex material properties in systems with pronounced relativistic effects and magnetic frustration. Spin-orbit coupling is a relativistic effect that gives rise to an interaction that couples an electron's spin and its orbital motion. SOC plays a critical role in stabilizing noncollinear magnetic structures in magnetic materials with heavy elements where the magnetic moments deviate from simple parallel or antiparallel alignments. This interplay is crucial for understanding exotic magnetic behaviors involving spin textures like skyrmions and spin-spirals. Spin-orbit coupling induces exotic electronic behaviors even in non-magnetic materials by lifting degeneracies in the electronic band structure, leading to phenomena such as band splitting and the stabilization of topological phases. The utilization of pseudopotential Density Functional Theory (DFT), a widely adopted first-principles material simulation tool, has been shown to be effective in predicting various material properties when extended to account for spin-orbit coupling~\cite{Kleinman1980RelativisticPseudopotential, Theurich2001Self-consistentPseudopotentials}.

Large-scale DFT calculations accounting for noncollinear magnetism and SOC are essential for accurately capturing complex physical phenomena, as demonstrated in several recent studies. For instance, \citet{Chatratin2023DopingCdTe} modeled dilute dopants in the semiconductor CdTe using supercells of up to 512 atoms, extrapolating their findings to the dilute limit which may not always be accurate. Similarly, \citet{Choi2016DensityMnSi} employed DFT with 512-atom supercells to investigate skyrmion pinning due to atomic defects in MnSi, noting finite-size effects in their results since the skyrmion size in MnSi ($\sim$ 10 nm) exceeds their supercell size (~3.4 nm). In another recent work, \citet{He2021GiantWTe2} used DFT to study the nonlinear Hall effect in incommensurate Moiré lattices of twisted bilayer WTe$_2$, simulating systems of up to 1032 atoms for twist angles as low as 12.3° and could not simulate systems with much lower twist angles. In all these cases, size limitations that can be handled in current DFT implementations, particularly when incorporating SOC and noncollinear magnetism, restricted the ability of these authors to study the desired physics from a first-principles perspective.

Among the various discretization methods employed for computations incorporating noncollinear magnetism and spin-orbit coupling effects in the framework of pseudopotential DFT, the plane-wave (PW) basis is the most widely utilized for solid-state systems, primarily due to its spectral convergence properties, which facilitate computationally efficient calculations. However, these plane-wave based methods suffer from well-known limitations, including poor scalability on parallel computing architectures, which limits their applications to material systems with at most a few thousand electrons. Plane-wave based methods are also inefficient for treating non-periodic systems due to their inability to handle generic boundary conditions. To address these challenges, there has been an increased focus on developing systematically convergent, efficient, and scalable real-space discretization techniques based on finite-difference~\cite{Kronik2006PARSECNanostructures,Michaud-Rioux2016RESCU:Method,Ghosh2017SPARC:Clusters,Ghosh2017SPARC:Systems,Andrade2015Real-spaceSystems,Enkovaara2010ElectronicMethod}, finite-elements~\cite{Tsuchida1995Electronic-structureMethod,Tsuchida1996AdaptiveCalculations,Pask1999Real-spaceApproach,Pask2005FiniteCalculations,Bylaska2009AdaptiveTheory,Suryanarayana2010Non-periodicTheory,Motamarri2013Higher-orderTheory,Schauer2013All-electronSpaces,Chen2014AdaptiveModels,Davydov2016OnImplementation,Kanungo2017Large-scaleBasis,Kanungo2019RealMethods,Motamarri2020DFT-FECalculations,Das2019FastSystem,Certik2024High-orderCalculations}, wavelets~\cite{Genovese2008DaubechiesCalculations,Genovese2009DensityArchitectures,Mohr2014DaubechiesTheory} and other reduced-order basis techniques~\cite{Hu2015DGDFT:Calculations,Motamarri2016Tucker-tensorCalculations,Lin2021Tensor-structuredCalculations,Lin2023TTDFT:Calculations}. However, we note that these systematically improvable real-space discretization strategies used in pseudopotential DFT calculations for noncollinear magnetism and spin-orbit coupling are predominantly based on finite-difference methods~\cite{Naveh2007Real-spaceTheory,Naveh2009Real-spaceTheory,Zhang2024SPARCFunctionals,Lu2021ImplementationCode}. These strategies have been applied to perform noncollinear DFT calculations on material systems containing up to a maximum of a few thousand electrons.

The finite-element (FE) basis, a relatively new entrant for real-space DFT calculations, is employed in the current work. These basis sets involve compactly supported piecewise polynomials and have emerged as a promising alternative, providing numerous advantages over other commonly used basis sets for density functional theory calculations. In particular, these basis functions can accommodate generic boundary conditions (periodic, semi-periodic, and non-periodic) while providing systematic convergence to the material properties of interest. Most importantly, the locality of these FE basis functions allows the development of computational algorithms that can exploit fine-grained parallelism on modern heterogeneous architectures~\cite{Das2023Large-ScaleAlloys,Das2022DFT-FEDiscretization,Panigrahi2024FastSystems}, ensuring excellent parallel scalability on distributed systems. Previous studies have shown that FE-based methods can significantly outperform plane-wave methods employing norm-conserving pseudopotential DFT calculations~\cite{Das2019FastSystem,Motamarri2020DFT-FECalculations} and very recently using the projector augmented method~\cite{Ramakrishnan2024FastMethod} for system sizes greater than 5000 electrons. The open-source DFT-FE code~\cite{Das2022DFT-FEDiscretization} inherits these features and incorporates efficient, scalable solvers for the Kohn-Sham DFT equations involving norm-conserving pseudopotentials within the framework of collinear magnetism. Furthermore, DFT-FE has demonstrated exceptional scalability on massively parallel many-core CPU and hybrid CPU-GPU architectures, handling simulations of material systems with up to 600,000 electrons on up to 200,000 CPU cores and 40,000 GPUs. The finite-element based methods incorporated in DFT-FE have also been used in various scientific studies recently to conduct large-scale DFT calculations involving material systems with tens of thousands of electrons~\cite{Zhuravel2020BackboneDNA,Menon2024AtomisticTemperatures,Kumar2023EffectAssessment,Shojaei2024Bridging/math,Das2019FastSystem,Das2023Large-ScaleAlloys,Yao2022ModulatingNanofilms,Ghosh2019All-electronSolids,Ghosh2021SpinspinCalculations,Kanungo2019ExactDensities}. 

We note that the aforementioned finite-element-based DFT pseudopotential calculations are all limited to the case of collinear magnetism and further do not incorporate spin-orbit effects. This work introduces a local real-space formalism incorporating noncollinear magnetism and spin-orbit coupling (SOC) within the norm-conserving pseudopotential DFT framework amenable for finite-element (FE) discretization building on the DFT-FE computational framework\cite{Das2022DFT-FEDiscretization,Motamarri2018ConfigurationalTheory}. Furthermore, we propose robust and efficient scalable strategies to evaluate the FE-discretized Hamiltonian and solve the underlying nonlinear generalized eigenvalue problem using a self-consistent field iteration approach. Additionally, we demonstrate that our method can handle fully periodic, non-periodic, and semi-periodic boundary conditions for generic material systems as large as 20,000 electrons within the framework of noncollinear magnetism involving 2-component complex spinors. Moreover, our method offers substantial computational efficiencies over state-of-the-art plane-wave-based approaches for large-scale systems, as demonstrated in this work. Notably, to our knowledge, this work represents the first development of a fast and scalable real-space computational approach incorporating noncollinear magnetism and spin-orbit coupling within the framework of finite-element discretization. We also introduce a generalized force approach to derive atomic forces and stresses within the DFT framework of noncollinear magnetism and SOC.

We adopt the locally collinear approximation to employ the existing approximate exchange-correlation functional forms that are well-tested for spin-collinear cases. The inclusion of spin-orbit interaction in the optimized norm-conserving pseudopotential (ONCV) framework employed in the current work is along the similar lines of \citet{Corso2005Spin-orbitPt}. The key aspects of the current work include: (i) utilizing the local real-space reformulation of the DFT electrostatics~\cite{Das2022DFT-FEDiscretization} to derive the governing equations in terms of the finite-element (FE) discretized  Hamiltonian matrix and the 2-component complex spinors to compute the ground-state magnetization and energies, (ii) devising a numerical strategy to evaluate the GGA exchange-correlation potentials in these governing equations, avoiding the computation of ill-defined gradients of magnetization direction near zero magnetization arising within the FE framework using $C^{0}$ basis functions, (iii) formulating a unified approach to compute atomic forces and unit-cell stresses by evaluating the directional derivatives of a generalized energy functional proposed in this work, extending the ideas of \citet{Methfessel1995IndependentMethods,Jacobsen1987InteratomicTheory} to the case of noncollinear magnetism, (iv) developing efficient and scalable methods to build the FE discretized Hamiltonian and thereby compute the action of this Hamiltonian on a trial subspace of vectors efficiently by exploiting the sparsity of local and non-local parts of this Hamiltonian, (v) employing a self-consistent iteration approach based on a Chebyshev filtered subspace iteration procedure that leverages these efficient strategies to solve the underlying nonlinear FE discretized generalized eigenvalue problem. The proposed formulation has been implemented in a distributed environment using both CPUs and GPUs, utilizing the message-passing interface (MPI) for communication across multiple nodes, enabling large-scale DFT calculations that account for noncollinear magnetism and spin-orbit effects. We benchmark the accuracy and performance of our method against plane-wave-based implementations on various representative non-periodic, semi-periodic, and fully periodic systems. Compared to reference data from plane-wave calculations, our results show excellent agreement in ground-state energies, band structures, vertical ionization potentials, magnetic anisotropy energies, and spin textures. Relative to widely used plane-wave based implementations, our finite-element based approach demonstrates significant computational advantage both in terms of CPU node-hrs and minimum wall time with increasing system sizes.

The remaining sections of the manuscript are organized as follows. \Cref{sec:mathform} briefly introduces the mathematical formulation to account for noncollinear magnetism and spin-orbit coupling using 2-component spinors within norm-conserving pseudopotential DFT formalism and finally discusses the necessary governing equations to be solved utilizing a local reformulation of electrostatics. Subsequently, \cref{sec:computemethod} introduces the finite-element discretization for the governing differential equations and details the proposed numerical strategy for evaluating the gradient-corrected exchange-correlation potentials under the locally collinear approximation. \Cref{sec:implem} details the efficient and scalable solution procedure for solving the FE discretized generalized eigenvalue problem using a self-consistent field iteration approach and the Chebyshev filtered subspace iteration method. \Cref{sec:benchmarks} presents comprehensively the accuracy and performance benchmarks of our implementation compared with state-of-the-art plane-wave (PW) codes on representative benchmark systems and large-scale systems. We find that the results obtained from our methodology are well within the bound of chemical accuracy when compared to the state-of-the-art PW codes for a wide range of benchmarks, including total energy, magnetic anisotropy energy, vertical ionization potentials, and band-structures. We also demonstrate up to a 2x reduction in computational cost for systems with $\sim$15,000-20,000 electrons and $\sim$8x-11x speed-ups for semi-periodic and non-periodic systems in terms of minimum wall times compared to widely used plane-wave implementations on CPUs. Furthermore, we also demonstrate the significant computational advantage of our proposed methodology on GPUs. Finally, we conclude with a short discussion summarizing the key findings and outline future prospects arising from this work in \cref{sec:concl}

\section{\label{sec:mathform} Mathematical Formulation}
This section will briefly outline the mathematical formulation necessary to account for noncollinear magnetism and spin-orbit coupling within the framework of real-space pseudopotential density functional theory amenable for finite-element discretization in order to compute ground-state energies, forces and periodic unit-cell stresses.
\subsection{Spinor representation of the wavefunctions}
Noncollinear magnetism is incorporated in density functional theory by representing the single-electron wavefunctions as 2-component spinors ($\bvPsi_n$) ~\cite{Kubler1988DensityMagnetism,Barth1972AI} given by
\begin{align}
    &{\bvPsi}_n\left(\br\right)=\begin{bmatrix}\psi^{\uparrow}_n\left(\br\right) \\ \psi^{\downarrow}_n\left(\br\right)\end{bmatrix} \forall n=1,\dots, N_e
\end{align}
where $N_e$ denotes the number of electrons in the given material system. The $2\times 2$ Hermitian density matrix $\bvrho$ in the spin space can then be expressed as  
\begin{align}
    \bvrho\left(\br\right)=\begin{bmatrix}\rho^{\uparrow\uparrow}\left(\br\right) & \rho^{\uparrow\downarrow}\left(\br\right) \\ \rho^{\downarrow\uparrow}\left(\br\right) & \rho^{\downarrow\downarrow}\left(\br\right)\end{bmatrix}&&
    \varrho^{\alpha\beta}\left(\br\right)&=\sum_{n}f_n{\psi^\alpha_n}^*\psi^{\beta}_n
\end{align}
where $\alpha$ and $\beta$ can be either $\uparrow$ (0) or $\downarrow$ (1) and $*$ denotes the complex conjugate. The above density matrix in the basis of the $2\times 2$ identity matrix ($\bI$) and the Pauli vector ($\vec\bsigma=\bsigma^x\hat\bx+\bsigma^y\hat\by+\bsigma^z\hat\bz$) can be written as
\begin{align}
\bvrho\left(\br\right)&=\frac{1}{2}\left(\rho\left(\br\right)\bI+\bmm\left(\br\right)\cdot\vec\bsigma\right)
\end{align}
Here $\rho\left(\br\right)$, the electron density and $\bmm\left(\br\right)$, the magnetization density are given by
\begin{align}    
\rho\left(\br\right)&=\sum_{n}f_n\bvPsi_n^\dagger\left(\br\right)\bvPsi_n\left(\br\right)\label{eqn:rho}\\
    \bmm\left(\br\right)&=\sum_nf_n\bvPsi_n^\dagger\left(\br\right)\vec\bsigma\bvPsi_n\left(\br\right)\label{eqn:mag}
\end{align}
where $f_n$ is the Fermi occupation number corresponding to $\bvPsi_n$ and $\dagger$ denotes the complex conjugate transpose.
\subsection{Kohn-Sham energy functional and the governing equations}
The Kohn-Sham energy functional associated with a material system comprising $N_a$ nuclei and $N_e$ electrons in the norm-conserving pseudopotential framework under the generalized gradient approximation (GGA) is written in terms of the 2-component spinors $\bPsi=\left[\bvPsi_1,\bvPsi_2,\dots,\bvPsi_{N}\right]$ with $N\geq N_e$, the vector of atomic coordinates $\bR=\left[\bR_1,\bR_2,\dots,\bR_{N_a}\right]$, the charge density $\rho$, and the magnetization density $\bmm$,  as follows:~\cite{Kubler1988DensityMagnetism}
\begin{multline}
E\left[\bPsi,\bR\right]=T_\text{s}\left[\bPsi\right]+E_\text{el}\left[\rho,\bR\right]+E_\text{psp}\left[\bPsi,\bR\right]\\
+E_\text{xc}\left[\rho,\bmm,\nabla\rho,\nabla\bmm\right]\label{eqn:functional}
\end{multline}
where the kinetic energy of the non-interacting electrons $\left(T_\text{s}\left[\bPsi\right]\right)$ is given by 
\begin{align}
    T_\text{s}[\bPsi]&=\sum_nf_n\int \left(\frac{\nabla\bvPsi_n^\dagger\left(\br\right)\cdot\nabla\bvPsi_n\left(\br\right)}{2}\right)d\br\label{eqn:kinetic}
\end{align}
while the total electrostatic energy $\left(E_\text{el}\left[\rho,\bR\right]\right)$ is  evaluated by defining the total nuclear charge density $b\left(\br,\bR\right) = \sum_a b^a\left(\br-\bR_a\right)$ with atom-centered smeared nuclear charges for each of the $N_a$ atoms is defined as $b^a\left(\br-\bR_a\right) \; \forall a=1,\dots,N_a$ ~\cite{Das2022DFT-FEDiscretization}. To this end, we have
\begin{multline}
E_\text{el}\left[\rho,\bR\right]=\frac{1}{2}\int\left(\rho(\br)+b(\br,\bR)\right)V_\text{el}(\br)\dr\\-\frac{1}{2}\sum_a\int b^a(\br^\prime-\bR_a)V_\text{self}^a(\br^\prime,\bR_a)\dr^\prime\label{eqn:electrostaticEnergy}
\end{multline}
wherein the total electrostatic potential, $V_\text{el}\left(\br,\bR\right)$, and the nuclear self-interaction potential for atom $a$, $V^a_\text{self}\left(\br,\bR_a\right)$, are obtained as the solutions of the following Poisson equations
\begin{align}
&-\nabla^2V_\text{el}\left(\br\right)=4\pi\left(\rho\left(\br\right)+b\left(\br,\bR\right)\right)\label{eqn:poissonEl}\\
&-\nabla^2V^{a}_\text{self}\left(\br,\bR_a\right)=4\pi b^a\left(\br-\bR_a\right)
\end{align}
and for the exchange-correlation energy $\left(E_\text{xc}\left[\rho,\bmm,\nabla\rho,\nabla\bmm\right]\right)$, we use the locally collinear approximation in order to utilize the existing approximate exchange-correlation functionals which have been well tested for the spin-collinear systems and are usually of the form ~\cite{Lehtola2018RecentTheory}
\begin{align}
    E_\text{xc}=\int f_{\text{xc}}(\rho^\uparrow,\rho^\downarrow,\gamma_0,\gamma_1,\gamma_2) \dr\label{eqn:xcEnergy}
\end{align}
where $f_{\text{xc}}(\rho^\uparrow,\rho^\downarrow,\gamma_0,\gamma_1,\gamma_2)$ is the exchange-correlation energy density. To this end, we make the following substitution for the spin-up ($ \rho^\uparrow$) and spin-down ($\rho^\downarrow$) charge densities.
\begin{align*}
    \rho^\uparrow&=\frac{\rho+\abs{\bmm}}{2} &  \rho^\downarrow&=\frac{\rho-\abs{\bmm}}{2}
\end{align*}
and the auxiliary quantities typically considered for the gradient-type exchange-correlation functionals are evaluated as
\begin{align*}
   \gamma_0 &=\frac{\nabla\left(\rho+\abs{\bmm}\right)\cdot \nabla \left(\rho+\abs{\bmm}\right)}{4}\\
    \gamma_1 &=\frac{\nabla\left(\rho+\abs{\bmm}\right)\cdot \nabla\left(\rho-\abs{\bmm}\right)}{4}\\
    \gamma_2&=\frac{\nabla\left(\rho-\abs{\bmm}\right)\cdot \nabla\left(\rho-\abs{\bmm}\right)}{4}
\end{align*}
We note that the above substitution is not unique, and various other substitutions have been proposed in literature ~\cite{Knopfle2000Spin-iron,Bulik2013NoncollinearProperties,Pu2023NoncollinearTheory} to enable the use of collinear-spin exchange-correlation functionals for the noncollinear spin case.

For the pseudopotential approximation we utilize the Optimized Norm-Conserving Vanderbilt (ONCV)~\cite{Hamann2013OptimizedPseudopotentials} pseudopotentials which allow for the following separable form for the pseudopotential contribution to the energy functional, $E_\text{psp}\left[\bPsi\right]=E_\text{loc}\left[\bPsi\right]+E_\text{nloc}\left[\bPsi\right]$. 
The local pseudopotential energy contribution, $E_\text{loc}\left[\bPsi,\right]$, can be evaluated as
\begin{align}
    E_\text{loc}\left[\bPsi,\bR\right]&=\int\left(V_\text{loc}(\br)-V_\text{self}(\br)\right)\rho(\br)d\br\label{eqn:lpspEnergy}
\end{align}
where the local pseudopotential operator can be written as the sum of the atom-dependent local pseudopotentials, $V_\text{loc}(\br)=\sum_aV_\text{loc}^a(\br - \bR_{a})$ and the nuclear self-interaction potential is given by $V_\text{self}(\br)=\sum_aV_\text{self}^a(\br,\bR_{a})$. Note that $V_\text{self}$ is subtracted here to account for the inclusion of the nuclear potential arising due to smeared charges ~\cite{Motamarri2020DFT-FECalculations,Das2022DFT-FEDiscretization} in the electrostatic energy term in \cref{eqn:electrostaticEnergy}

For the case of spin-orbit coupling in the ONCV framework~\cite{Corso2005Spin-orbitPt}, the non-local pseudopotential contribution to the energy functional, $E_\text{nloc}\left[\bPsi\right]$, can be expressed as
\begin{align}    E_\text{nloc}\left[\bPsi\right]&=\sum_nf_n\int\int\bvPsi_n^\dagger(\br)\bV_{\!\text{nloc}}(\br,\br^\prime)\bvPsi_n(\br^\prime)\dr\dr^\prime\label{eqn:nlpspEnergy}
\end{align}
and the non-local pseudopotential operator, $\bV_{\!\text{nloc}}(\br,\br^\prime)$ is given by
\begin{align}
    \bV_{\!\text{nloc}}(\br,\br^\prime)=\sum_{a}\sum_{\rchi\rchi^\prime}\bD^{\gamma_a,\rchi,\rchi^\prime}p^a_\rchi(\br-\bR_a)p^a_{\rchi^\prime}(\br^\prime-\bR_a)\label{eqn:nlOP}
\end{align}
where $\gamma_a$ denotes the atom type of atom $a$ and we define composite indices $\rchi=\left\{\tau,l,j,m\right\}$ and $\rchi^\prime=\left\{\tau^\prime,l^\prime,j^\prime,m^\prime\right\}$ such that $l$ ($l^\prime$) and $j$ ($j^\prime$) denote the orbital and the total angular momentum respectively, $m$ ($m^\prime$) denotes the projection of the angular momentum on the quantization axis. Further, $p^a_{\rchi}(\br)$ is the non-local projector function indexed by $\tau$ centered at atom $a$, and the angular momentum components $l$, $j$ and $m$, while $\bD^{\gamma_a,\rchi,\rchi^\prime}$ are the $2\times2$ matrices representing the non-local pseudopotential coefficients.

The Euler-Lagrange equations corresponding to minimizing the energy functional in~\cref{eqn:functional} subject to the orthogonality constraint $\int\bvPsi_n^\dagger(\br)\bvPsi_{n^\prime}(\br)\dr=\delta_{nn^\prime}$ can be written as the following nonlinear Hermitian eigenvalue problem 
\begin{align}
    \mathcal{H}\bvPsi_n=\epsilon_n\bvPsi_n\label{eqn:eigenProb}
\end{align}
to be solved for eigenfunctions corresponding to $N$ smallest eigenvalues,
where $\{\epsilon_n\}$ are the eigenvalues and $\{\bvPsi_n\}$ are the corresponding eigenfunctions of the Hamiltonian operator $\mathcal{H}$ which are the canonical wavefunctions that minimize the energy functional~\cref{eqn:functional}. We note that $\mathcal{H}$ can be decomposed as $\mathcal{H}=\mathcal{H}^\text{loc}+\mathcal{H}^\text{nloc}$ with $\mathcal{H}^\text{loc}$ defined by
\begin{align}
    \mathcal{H}^\text{loc}=\left[-\frac{1}{2}\nabla^2+V_\text{eff}(\br)\right]\bI+\bB_\text{xc}\cdot\vec\bsigma
\end{align}
where $\bI$ is the $2 \times 2$ identity matrix and $\vec\bsigma$ is the Pauli vector. The effective potential and the XC fields are defined as
\begin{align}
    V_\text{eff}(\br)=V_\text{el}(\br)+V_\text{xc}(\br)+(V_\text{loc}(\br)-V_\text{self}(\br))
\end{align}
\begin{align}
    V_\text{xc}(\br)&=\frac{\delta E_\text{xc}}{\delta\rho(\br)}&\bB_\text{xc}&=\frac{\delta E_\text{xc}}{\delta\bmm(\br)}
\end{align}
and $\mathcal{H}^\text{nloc}$ is defined as follows
\begin{align}
        &\mathcal{H}^\text{nloc}\bvPsi_n:=\int \bV_{\!\text{nloc}}(\br,\br^\prime)\bvPsi_n(\br^\prime)d\br^\prime
\end{align}
Using the separable form of the non-local pseudopotential operator from \cref{eqn:nlOP}, we have
\begin{multline}
\int \bV_{\!\text{nloc}}(\br,\br^\prime)\bvPsi_n(\br^\prime)d\br^\prime 
=\sum_a\sum_{\rchi\rchi^\prime}p_\rchi^a(\br-\bR_a)\bD^{\gamma_a,\rchi,\rchi^\prime}\\\int p^a_{\rchi^\prime}(\br^\prime-\bR_a)\bvPsi_n(\br^\prime)d\br^\prime
\end{multline}
All the integrals in \cref{eqn:electrostaticEnergy,eqn:kinetic,eqn:kinetic,eqn:xcEnergy,eqn:lpspEnergy,eqn:nlpspEnergy} are over the entire space ($\mathbb{R}^{3}$) for the case of non-periodic systems. For the case of periodic systems, the energy functional in \cref{eqn:functional} represents the energy per the periodic unit cell and all the integrals involving $\br$ in \cref{eqn:electrostaticEnergy,eqn:kinetic,eqn:kinetic,eqn:xcEnergy,eqn:lpspEnergy,eqn:nlpspEnergy} are over this unit-cell and those involving $\br^\prime$ are over the entire space $\mathbb{R}^{3}$. Furthermore, we can now invoke the Bloch theorem, $\bvPsi_n(\br)=e^{\iu \bk\cdot\br}\bu_{n\bk}(\br)$ with $\bu_{n\bk}(\br)$ denoting the 2-component complex-valued periodic Bloch wavefunctions and consequently, the summations over the eigenvector index accounting for the Brillouin zone integration can be written as 
\begin{align}
    \sum_nf_n \rightarrow \sum_n\fint_{BZ}f_{n\bk}\dk
\end{align}
where $\fint_{BZ}$ denotes the volume average of the integral over the Brillouin zone corresponding to the periodic unit-cell $\Omega_p$ and $f_{n\bk}$ are the orbital occupation numbers corresponding to $\bu_{n\bk}$. To this end, the nonlinear eigenvalue problem in \cref{eqn:eigenProb} can be recast as follows:
\begin{align}   \mathcal{H}^{\bk}\bu_{n\bk}=\epsilon_{n\bk}\bu_{n\bk}\label{eqn:eigenProbPer}
\end{align}
where the transformed $\bk$-dependent Hamiltonian operator is defined as $\mathcal{H}^{\bk}=e^{-\iu\bk\cdot\br}\mathcal{H}e^{\iu\bk\cdot\br}=\mathcal{H}^{\bk,loc}+\mathcal{H}^{\bk,nloc}$. Here, $\mathcal{H}^{\bk,loc}$ is given by
\begin{multline}
    \mathcal{H}^{\bk,loc}=\left[-\frac{1}{2}\nabla^2-\iu \bk\cdot\nabla+\frac{\abs{\bk}^2}{2}+V_\text{eff}(\br)\right]\bI\\+\frac{\delta E_\text{xc}}{\delta\bmm(\br)}\cdot\bsigma
\end{multline}
and $\mathcal{H}^{\bk,nloc}$ is given by
\begin{multline}
    \mathcal{H}^{\bk,nloc} \bu_{n\bk}=\sum_{a\in \Omega_p}\sum_{\rchi \rchi^\prime}\sum_q e^{-\iu\bk\cdot(\br-\bL_q)}p_\rchi^a(\br-\bL_q-\bR_a)\\\bD^{\gamma_a,\rchi,\rchi^\prime}\int_{\Omega_p} \sum_{q^\prime}e^{\iu\bk\cdot(\br^\prime-\bL_{q^\prime})}p^a_{\rchi^\prime}(\br^\prime-\bL_{q^\prime}-\bR_a)\bu_{n\bk}(\br^\prime)d\br^\prime\label{eqn:HnlocPeriodic}
\end{multline}
where $\bL_q$ and $\bL_{q^\prime}$ are the lattice vectors.

To summarize, the governing equations to be solved for density functional theory incorporating noncollinear magnetism and spin-orbit coupling are given by
\begin{widetext}
\begin{equation}
    \boxed{    \begin{gathered}
    \mathcal{H}^{\bk}\bu_{n\bk}=\epsilon_{n\bk}\bu_{n\bk}\\
    \mathcal{H}^{\bk}=\mathcal{H}^{\bk,loc}+\mathcal{H}^{\bk,nloc}\\
    \mathcal{H}^{\bk,loc}=\left[-\frac{1}{2}\nabla^2-\iu \bk\cdot\nabla+\frac{\abs{\bk}^2}{2}+V_\text{eff}(\br)\right]\bI+\bB_\text{xc}\cdot\bsigma\\
    \mathcal{H}^{\bk,nloc} \bu_{n\bk}=\sum_{a\in \Omega_p}\sum_{\rchi \rchi^\prime}\sum_{q} e^{-\iu\bk\cdot(\br-\bL_q)}p_\rchi^a(\br-\bL_q-\bR_a)\bD^{\gamma_a,\rchi,\rchi^\prime}\int_{\Omega_p} \sum_{q^\prime}e^{\iu\bk\cdot(\br^\prime-\bL_{q^\prime})}p^a_{\rchi^\prime}(\br^\prime-\bL_{q^\prime}-\bR_a)\bu_{n\bk}(\br^\prime)d\br^\prime\\
        V_\text{eff}(\br)=V_\text{el}(\br)+V_\text{xc}(\br)+(V_\text{loc}(\br)-V_\text{self}(\br))\\
        V_\text{loc}(\br)=\sum_q\sum_{a\in\Omega_p}V_\text{loc}^a(\br - \bL_q -\bR_{a})\qquad
        V_\text{self}(\br)=\sum_q\sum_{a\in\Omega_p}V_\text{self}^a(\br, \bL_q +\bR_{a})\\
        V_\text{xc}(\br)=\frac{\delta E_\text{xc}}{\delta\rho(\br)}\qquad\bB_\text{xc}=\frac{\delta E_\text{xc}}{\delta\bmm(\br)}\\
        -\nabla^2V_\text{el}\left(\br\right)=4\pi\left(\rho\left(\br\right)+b\left(\br,\bR\right)\right)\\
        -\nabla^2V^{a}_\text{self}\left(\abs{\br-\bR_a}\right)=4\pi b^a\left(\abs{\br-\bR_a}\right)
    \end{gathered}}\label{eqn:nlEig}
\end{equation}
\end{widetext}
This nonlinear eigenvalue problem can be formulated as a fixed point iteration problem as
\begin{align}
    F\left[\left(\rho,\bmm\right)\right]=\left(\rho,\bmm\right)\label{eqn:KSMap}
\end{align}
where the map $F[\left(\rho_{in},\bmm_{in}\right)]=\left(\rho_{out},\bmm_{out}\right)$ represents the computation of $V_\text{eff}^{in}$ and $\bB_\text{xc}^{in}$ using a guess of input densites $\left(\rho_{in},\bmm_{in}\right)$, solving eigenvalue problem given by \cref{eqn:nlEig} and computing $\left(\rho_{out},\bmm_{out}\right)$ using \cref{eqn:rho,eqn:mag}.

Upon obtaining the solution of the above fixed-point iteration to a required tolerance on $\norm{\rho_{out}-\rho_{in}}$ and $\norm{\bmm_{out}-\bmm_{in}}$, the total free energy (per unit-cell in case of a periodic system) can be obtained through the double-counting method~\cite{Martin2020ElectronicStructure} as
\begin{multline}
    E_{0}=\sum_{n}\fint_{BZ}f_{n\bk}\epsilon_{n\bk}\dk-\int \left(V_\text{elxc}^{in}\rho_{out}+\bB_\text{xc}^{in}\cdot\bmm_{out}\right)\dr\\
    +E_\text{el}[\rho_\text{out}]+E_\text{xc}[\rho_\text{out},\bmm_\text{out}]+E_\text{ent}[f_{n\bk}]\label{eqn:doubleCountingEnergy}
\end{multline}
Where $V_\text{elxc}=V_\text{el}+V_\text{xc}$ and the entropic energy contribution ($-TS$ where $T$ is the temperature and $S$ is the entropy), $E_\text{ent}$, is given by
\begin{multline}
    E_\text{ent}=k_BT\sum_n\fint_{BZ}f_{n\bk}\ln{\left(f_{n\bk}\right)}\dk\\+k_BT\sum_n\fint_{BZ}\left(1-f_{n\bk}\right)\ln{\left(1-f_{n\bk}\right)}\dk
\end{multline}
We will now discuss the evaluation of the derivatives of energy, specifically the atomic forces and unit-cell stresses.
\subsection{Derivatives of energy: Atomic forces and cell-stresses}
In order to compute the derivative of energy, we make use of the configurational force approach previously used by \citet{Motamarri2018ConfigurationalTheory,Rufus2022IonicBasis,Das2022DFT-FEDiscretization} in order to compute the derivatives for the spin-collinear and spin unpolarized cases. To this end, the starting point of these works to evaluate the configurational force is to consider the Kohn-Sham energy functional with the non-orthogonal electronic wavefunctions and the single-particle density matrix as independent fields. This makes the derivation of configurational forces non-trivial and makes it even more challenging to account for two-component spinors and the 2 $\times$ 2 density matrix arising in noncollinear magnetism.
Consequently, in this work, we introduce a generalized energy functional in the spirit of energy expression given in \cref{eqn:doubleCountingEnergy} as the starting point of our derivation, extending the ideas of \citet{Methfessel1995IndependentMethods,Jacobsen1987InteratomicTheory} to the case of noncollinear magnetism. This energy functional is given by
\begin{multline}
    \widetilde{E}[\widetilde V,\widetilde \bB,\widetilde \rho,\widetilde \bmm,\widetilde f_{n\bk}]=\sum_{n}\fint_{BZ}\widetilde f_{n\bk}\epsilon_{n\bk}[\widetilde V,\widetilde\bB]\dk\\-\int \left(\widetilde V\widetilde\rho+\widetilde\bB\cdot\widetilde\bmm\right)\dr
    +E_\text{el}[\widetilde\rho]+E_\text{xc}[\widetilde\rho,\widetilde\bmm]\\+E_\text{ent}[\widetilde f_{n\bk}]+\mu\left(N_e-\sum_{n}\fint_{BZ}\widetilde{f}_{n\bk}\dk\right)\label{eqn:genEnergyFunc}
\end{multline}
and has the constraint on the number of electrons imposed -via- the Lagrange multiplier $\mu$. In the above functional~\ref{eqn:genEnergyFunc}, we treat the potential $\widetilde{V}$, the magnetic field $\widetilde\bB$, the total charge density $\widetilde\rho$, the magnetization density vector $\widetilde\bmm$ and the occupation numbers $\widetilde f_{n\bk}$ as the variational parameters. Note that here, $\epsilon_{n\bk}\left[\widetilde V,\widetilde\bB\right]$ are defined as the eigenvalues corresponding to the eigenvalue problem defined by
\begin{gather}
    \mathcal{H}^{\bk}\bu_{n\bk}=\epsilon_{n\bk}\bu_{n\bk}\nonumber\\
    \mathcal{H}^{\bk}=\mathcal{H}^{\bk,loc}+\mathcal{H}^{\bk,nloc}\nonumber\\
    \mathcal{H}^{\bk,loc}=\left[-\frac{1}{2}\nabla^2-\iu \bk\cdot\nabla+\frac{\abs{\bk}^2}{2}+V_\text{eff}(\br)\right]\bI+\widetilde\bB\cdot\bsigma\nonumber\\
     V_\text{eff}(\br)=\widetilde V+(V_\text{loc}(\br)-V_\text{self}(\br))\label{eqn:genKSProblem}
\end{gather}
with $\mathcal{H}^{\bk,nloc}$ defined by \cref{eqn:HnlocPeriodic}
This functional in ~\cref{eqn:genEnergyFunc} has a stationary point at the minima corresponding to Kohn-Sham DFT energy functional (see \cref{sec:appendixGenFunc} for the proof). Let $\widetilde{E}_S=\widetilde{E}[\widetilde V=V_\text{elxc},\widetilde \bB=\bB_\text{xc},\widetilde \rho=\rho,\widetilde \bmm=\bmm,\widetilde f_{n\bk}=f_{n\bk}]=E_0$ denote the stationary point of this functional.

We now consider a parametrized perturbation of the underlying space described by $\btau^{\varepsilon}$ which maps a point $\br$ in the unperturbed space to a point $\br^\varepsilon=\btau^{\varepsilon}(\br)$ in the perturbed space. We also define the generator of this perturbation as $\bUpsilon=\frac{d\btau^\varepsilon}{d\varepsilon}\bigl|_{\varepsilon=0}$, in this framework and we can compute the configurational force due to the pertubation $\btau^{\varepsilon}$ by evaluating the following directional derivative (see \cref{sec:appendixGatDer} for the derivation of the generalized force expression).
\begin{widetext}
    \begin{align}
        \frac{d\widetilde{E}_S(\btau^{\varepsilon})}{d\varepsilon}\biggl|_{\varepsilon=0}=\frac{d\widetilde{E}_S^{\varepsilon}}{d\varepsilon}\biggl|_{\varepsilon=0}=\int_{\Omega_p}\bE:\nabla\bUpsilon\dr+\sum_{a\in \Omega_p}\int_{\mathbb{R}^3}\bE^a:\nabla\bUpsilon\dr+\text{F}^\text{psp,nloc}+\text{F}^{K}+\text{F}^\text{ext,corr}+\text{F}^\text{sm}\label{eqn:GenForce}
    \end{align}
    where `$:$' denotes a tensor contraction and the rank-2 tensor $\bE^a$ is given by
    \begin{align}
        \bE^a&=\frac{1}{8\pi}\left(\nabla V_\text{self}^a\cdot\nabla V_\text{self}^a\right)\bI-\frac{1}{4\pi}\nabla V_\text{self}^a\otimes \nabla V_\text{self}^a
    \end{align}
    where `$\otimes$' and `$\cdot$' denote the outer and inner products over the spatial dimensions respectively.

    Similiarly the rank-2 tensor $\bE$ is given by
    \begin{multline}
        \bE=\Biggl(\sum_{n}\fint_{BZ} \frac{f_{n\bk}}{2}\left(\nabla\bu_{n\bk}^\dagger\cdot\nabla\bu_{n\bk}+\left(\abs{\bk}^2-\epsilon_{n\bk}\right)\bu_{n\bk}^\dagger \bu_{n\bk}-\iu \bu_{n\bk}^\dagger\bk\cdot\nabla \bu_{n\bk}\right)\dk+\left(V_\text{loc}-V_\text{self}\right)\rho+f_\text{xc}+\rho V_\text{el}-\frac{\abs{\nabla V_\text{el}}^2}{8\pi} \Biggr)\bI\\-\sum_{n}\fint_{BZ} \frac{f_{n\bk}}{2}\left(\nabla\bu_{n\bk}^\dagger\otimes\nabla\bu_{n\bk}+\nabla\bu_{n\bk}\otimes\nabla\bu_{n\bk}^\dagger-\iu\bu_{n\bk}^\dagger\nabla\bu_{n\bk}\otimes\bk\right)\dk+\frac{1}{4\pi}\nabla V_\text{el}\otimes\nabla V_\text{el}\\-\parder{f_\text{xc}}{\nabla\rho}\otimes\nabla\rho-\parder{f_\text{xc}}{\nabla \abs{\bmm}}\otimes\nabla \abs{\bmm}
    \end{multline}

    The term $\text{F}^\text{psp,nloc}$ is given by $\text{F}^\text{psp,nloc}=\text{F}^\dagger_\text{nloc}+\text{F}_\text{nloc}$ where $\text{F}_\text{nloc}$ is given by
    \begin{multline}
        \text{F}_\text{nloc}=\sum_{a\in \Omega_p}\sum_{\rchi \rchi^\prime}\sum_q  \sum_{n}\fint_{BZ}\Biggl[\int_{\Omega_p}\biggl(\bu_{n\bk}^\dagger(\br) e^{-\iu\bk\cdot(\br-\bL_q)}p_\rchi^a(\br-\bL_q-\bR_a)\biggr)\dr\bD^{\gamma_a,\rchi,\rchi^\prime}\int_{\Omega_p} \biggl(\sum_{q^\prime}e^{\iu\bk\cdot({\br^\prime}-\bL_{q^\prime})}\\p^a_{\rchi^\prime}({\br^\prime}-\bL_{q^\prime}-\bR_a)\left(-\left(\bUpsilon(\br^\prime)-\bUpsilon\left(\bR_a+\bL_{q^\prime}\right)\right)\cdot\nabla\bu_{n\bk}(\br^\prime)-\iu\bk\cdot\bUpsilon(\bR_a)\bu_{n\bk}(\br^\prime)\right)\biggr)\dr^\prime\Biggr]\dk
    \end{multline}
    The term $\text{F}^K$ is given by
    \begin{multline}
        \text{F}^K=\sum_{n}\fint_{BZ}\int_{\Omega_p}\Biggl[-\iu\bu_{n\bk}^\dagger\pardervepsZ{\bk^\varepsilon}\cdot\nabla\bu_{n\bk}+\frac{1}{2}\pardervepsbZ{\abs{\bk^\varepsilon}^2}\bu_{n\bk}^\dagger\bu_{n\bk}\Biggr]\dr\dk\\+\sum_{a\in \Omega_p}\sum_{\rchi \rchi^\prime}\sum_q \sum_{n}\parderveps{}\Biggl\{\fint_{BZ}\Biggl[\int_{\Omega_p}\biggl(\bu_{n\bk}^\dagger(\br) e^{-\iu\bk^\varepsilon\cdot(\br-\bL_q)}p_\rchi^a(\br-\bL_q-\bR_a)\biggr)\dr\\\bD^{\gamma_a,\rchi,\rchi^\prime}\int_{\Omega_p} \biggl(\sum_{q^\prime}e^{\iu\bk^\varepsilon\cdot({\br^\prime}-\bL_{q^\prime})}p^a_{\rchi^\prime}({\br^\prime}-\bL_{q^\prime}-\bR_a)\bu_{n\bk}(\br^\prime)\biggr)\dr^\prime\Biggr]\dk\Biggr\}\Bigg|_{\varepsilon=0}
    \end{multline}
    The term $\text{F}^\text{ext,corr}$ is given by
    \begin{multline}
        \text{F}^\text{ext,corr}=\sum_{a\in\Omega_p}\sum_q\int_{\Omega_p}\rho\Biggl(\nabla V_\text{loc}^a(\br-\bR_a-\bL_q)\cdot\left(\bUpsilon(\br)-\bUpsilon(\bR_a+\bL_q)\right)-\nabla V_\text{self}^a(\br,\bR_a+\bL_q)\cdot\bUpsilon(\br)\\-\pardervepsZ{V_\text{self}^a(\br,b^a({\br^\prime}^\varepsilon-\bR_a^\varepsilon-\bL_q^\varepsilon))}\Biggr)\dr
    \end{multline}
    Finally, the term $\text{F}^\text{sm}$ is given by
    \begin{align}
        \text{F}^\text{sm}&=\sum_{a\in \Omega_p}\sum_q\int_{\Omega_p}b^a(\br-\bR_a-\bL_q)\nabla V_\text{el}\cdot\left(\bUpsilon(\br)-\bUpsilon(\bR_a+\bL_q)\right)\dr-\sum_{a\in \Omega_p}\int_{\mathbb{R}^3}b^a(\br-\bR_a)\nabla V_\text{self}^a\cdot(\bUpsilon(\br)-\bUpsilon(\bR_a))\dr
    \end{align}
\end{widetext}

Now, the atomic forces and cell stress can be computed by choosing appropriate generators ($\bUpsilon$). The $i^{th}$ component of the force on atom $j$ can be computed using a generator whose $i^{th}$ component is compactly supported around atom $j$, the other components of the generator being zero. Cell stresses can be evaluated using a generator corresponding to an appropriate affine transformation. We refer to earlier works~\cite{Das2022DFT-FEDiscretization,Rufus2022IonicBasis,Motamarri2018ConfigurationalTheory} for further discussion on the choice of the generator and efficient evaluation of the terms arising in \cref{eqn:GenForce}.
\section{\label{sec:computemethod} Finite-element based Computational Methodology}
In this section, we will discuss the higher-order finite-element based numerical methodologies for solving the Kohn-Sham eigenvalue problems in \cref{eqn:eigenProb,eqn:eigenProbPer}. To this end,  we first provide a brief overview of the finite-element (FE) basis and describe the discretization of the underlying DFT governing equations involving noncollinear magnetism with spin-orbit coupling effects. Subsequently, we describe the computation of the effective potential terms in the discrete FE setting, followed by the numerical methodology to solve the resulting discretized nonlinear generalized eigenvalue problem. Furthermore, we provide a brief discussion of the efficient numerical implementation strategies underlying the proposed methodology that is well-suited for modern supercomputing architectures.

\subsection{Finite-element discretization}\label{sec:FEDiscretization}
In the finite-element (FE) method, the given spatial domain of interest is decomposed into non-overlapping subdomains called finite-elements (cells) by generating an FE mesh. The key aspect of this finite-element subspace is that the underlying basis functions are systematically convergent, compactly supported, piecewise $C^0$ continuous polynomials~\cite{Brenner2008TheMethods,Hughes2012TheAnalysis,Bathe2006FiniteProcedures,Ciarlet2002TheProblems}, amenable for massive parallelization. The advantages provided by higher-order adaptive spectral finite-element based methods to solve the Kohn-Sham DFT problem have been discussed in prior works~\cite{Motamarri2013Higher-orderTheory,Motamarri2020DFT-FECalculations,Das2022DFT-FEDiscretization,Kanungo2017Large-scaleBasis} in the context of norm-conserving pseudopotentials and all-electron DFT calculations within the framework of collinear magnetism.  

Within the framework of noncollinear magnetism, the finite-element discretization of the 2-component complex spinors is given by
\begin{align}
\bu_{n\bk}^{h,p}\left(\br\right)&=\sum_{I=0}^{M^{h,p}-1}\begin{bmatrix}      u^{I,\uparrow}_{n\bk} \\[0.1in] u^{I,\downarrow}_{n\bk}  \end{bmatrix}N^{h,p}_I(\br) \nonumber \\
&=\sum_{I=0}^{M^{h,p}-1}\bu_{n\bk}^IN^{h,p}_I(\br)\label{eqn:FEWfns}
\end{align}
where $\bu_{n\bk}^I$ are the 2-component linear combination coefficients associated with the discretized complex Bloch wavefunction $\bu_{n\bk}^{h,p}(\br)$ and $N^{h,p}_I(\br): 0 \leq I < M^{h,p}$ are the 3D tensor-structured FE polynomial basis constructed from 1D Lagrange polynomials of degree $p$ defined over Gauss Lobatto Legendre (GLL) nodal points~\cite{Brenner2008TheMethods}, generated using the nodes of the FE triangulation $\mathcal{T}^h$ with the characteristic mesh size denoted by $h$. Consequently, the FE discretization of the nonlinear eigenvalue problem corresponding to \cref{eqn:eigenProbPer}(or equivalently \cref{eqn:eigenProb}) results in an algebraic generalized Hermitian eigenvalue problem that can be written as 
\begin{align}   \bH^{\bk}\bU^{\bk}=\bM\bU^{\bk}\bLam^{\bk}\label{eqn:GHEP}
\end{align}
where $\bU^{\bk}$ is the matrix with $n^{th}$ column formed by the 2-component complex linear combination coefficients corresponding to spinor wavefunction of index $n$. Hence, the matrix entries $\text{U}^{\bk}_{2I+\alpha,n}$ correspond to $u^{I,\alpha}_{n\bk}$ as defined in \cref{eqn:FEWfns}. The discretized Hamiltonian matrix $\bH^{\bk}=\bH^{\bk,\text{loc}}+\bH^{\bk,\text{nloc}}$, and the FE basis overlap matrix $\bM$ are $2M^{h,p}\times 2M^{h,p}$ complex Hermitian matrices. The local part of this Hamiltonian, $\bH^{\bk,\text{loc}}$, can now be written as
\begin{multline}    \text{H}_{2I+\alpha,2J+\beta}^{\bk,\text{loc}}=\\\int_{\Omega_p}\Biggl(\biggl[\frac{1}{2}\nabla N^{h,p}_I(\br)\cdot\nabla N^{h,p}_J(\br)-\iu N^{h,p}_I(\br)\bk\cdot\nabla N^{h,p}_J(\br)\\+\frac{\abs{\bk}^2}{2}N^{h,p}_I(\br)N^{h,p}_J(\br)+ V_\text{eff}(\br)N^{h,p}_I(\br)N^{h,p}_J(\br)\biggr]\delta_{\alpha\beta}\\+\sum_{d=x,y,z}\biggl[B^d_\text{xc}(\br)N^{h,p}_I(\br)N^{h,p}_J(\br)\biggr]\sigma^d_{\alpha\beta}\Biggr)\dr\label{eqn:FEHloc}
\end{multline}
where $I,J$ are the FE basis function indices ranging from $0$ to $M^{h,p}-1$ and $\alpha,\beta$ are the spin indices taking values of $\{0,1\}$. The non-local part of the FE discretized Hamiltonian, $\bH^{\bk,\text{nloc}}$ can be written as
\begin{align}  \text{H}_{2I+\alpha,2J+\beta}^{\bk,\text{nloc}}&=\sum_{a}\sum_{\rchi\rchi^\prime}\text{P}^{a,\bk}_{I\rchi^\prime} \text{D}^{\gamma_a,\rchi,\rchi^\prime}_{\alpha\beta} {\text{P}^{a,\bk}_{J\rchi}}^*\label{eqn:FEHnloc}
\end{align}
where the $M^{h,p}\times n^a_{\text{pj}}$ matrices $\bPt^{a,\bk}$, with $n^a_{\text{pj}}$ denoting the number of non-local projectors for atom $a$,  are defined as
\begin{align}    
    \text{P}^{a,\bk}_{I\rchi}&=\int_{\Omega_p}\sum_qe^{-\iu\bk\cdot(\br-\bL_q)}p^a_{\rchi}(\br-\bL_q)N^{h,p}_I(\br)\dr\label{eqn:CMat}
\end{align}
Finally, the FE basis overlap matrix is defined by
\begin{align}    \text{M}_{2I+\alpha,2J+\beta}=\delta_{\alpha\beta}\int_{\Omega_p}N^{h,p}_I(\br)N^{h,p}_J(\br)\dr\label{eqn:OvlapMap}
\end{align}
To evaluate the above integrals, we consider partitioning of the simulation domain $\Omega_p$ into non-overlapping hexahedral cells, $\Omega^{(e)}$. Denoting $E$ to be the number of FE cells, we have $\Omega_p=\bigcup_{e=1}^{E}\Omega^{(e)}$, and further, within each cell $\Omega^{(e)}$, we employ $n_p^3$ three-dimensional (3D) Lagrange polynomial basis functions that are tensor products of  1D Lagrange polynomials of degree $p$ ($n_p=p+1$) constructed over Gauss Lobatto Legendre (GLL) nodal points, where $n_p$ denotes the number of these nodal points in each spatial direction~\cite{Brenner2008TheMethods}. This ensures that a given Lagrange polynomial basis function $N^{h,p}_I(\br)$ centered at a FE nodal point $I$ is non-zero in a given element $\Omega^{(e)}$ if and only if $I \in \Omega^{(e)}$. Consequently, the integrals over $\Omega_p$ in \cref{eqn:FEHloc,eqn:CMat,eqn:OvlapMap} are decomposed into integrals over individual cells $\Omega^{(e)}$ and Gauss Legendre quadrature rules are employed to evaluate these integrals over a reference cell $\left[-1,1\right]^3$ mapped from $\Omega^{(e)}$ as is done traditionally in finite-element based methods~\cite{Hughes2012TheAnalysis}. To this end, we have
\begin{align}   \int_{\Omega_p}\dr\rightarrow\sum_e\int_{\Omega^{(e)}}\dr\rightarrow\sum_e\sum_qw_qJ^{(e)}\biggr\rvert_{\br^{(e)}_q}\label{eqn:cellIntegrals}
\end{align}
where $J^{(e)}$ is the determinant of $\bJ^{(e)}$, the Jacobian of the map from the FE cell $\Omega^{(e)}$ to the reference cell,  and $w_q$ and $\br^{(e)}_q$ denote the Gauss-Legendre quadrature weights and sampling points constructed as a tensor product of 1D Gauss-Legendre quadrature rules of order $n_q$. The order of the quadrature rule is chosen such that the quadrature errors are of higher order than that of the finite-element discretization error incurred in the ground-state energies. 
\subsubsection{Evaluation of effective potential}
We now begin with a discussion on the evaluation of the exchange-correlation terms within the framework of noncollinear magnetism ($V_\text{xc}$ and $\bB_\text{xc}$) as it requires non-trivial considerations in the finite-element setting, especially for the case of GGA functionals as described below:
\paragraph{Evaluation of $V_\text{xc}$ and $B_\text{xc}$:}
Under the locally collinear approximation~\cite{Lehtola2018RecentTheory}, the functional derivatives of the exchange-correlation energy can be written as
\begin{align}
    V_{\text{xc}}&=f^+-\nabla\cdot\bg^+\label{eqn:vxcDiv}\\
    \bB_\text{xc}&=\hat\bmm\left(f^--\nabla\cdot\bg^-\right)\label{eqn:bxcDiv}
\end{align}
where $\hat{\bmm}$ is the direction vector of the magnetization density given by $\hat{\bmm}={\bmm}/{\abs{\bmm}}$. The scalar fields $f^\pm$ and the vector fields $\bg^\pm$ are given by
\begin{align}
    f^{\pm}&=\frac{1}{2}\left(\parder{f^\text{xc}}{\rho^\uparrow}\pm\parder{f^\text{xc}}{\rho^\downarrow}\right)\label{eqn:fpm}
\end{align}
\begin{align}
    \bg^{\pm}&=\parder{f^\text{xc}}{\gamma_0}\nabla\rho^\uparrow\pm \parder{f^\text{xc}}{\gamma_2}\nabla\rho^\downarrow+\frac{1}{2}\parder{f^\text{xc}}{\gamma_1}\left(\nabla\rho^\downarrow\pm\nabla\rho^\uparrow\right)\label{eqn:bgpm}
\end{align}
From \cref{eqn:vxcDiv,eqn:bxcDiv,eqn:fpm,eqn:bgpm}, we see that the evaluation of the exchange-correlation potentials requires the computation of Laplacians of the charge densities, which can lead to numerical difficulties~\cite{Martin2020ElectronicStructure}. Furthermore, these numerical issues are compounded by the fact that the $C^0$ continuous finite-element basis employed to discretize the electronic fields in the current work does not allow for the computation of higher-order derivatives accurately. However, as can be seen from \cref{eqn:FEHloc}, one only requires the evaluation of integrals involving exchange-correlation terms and hence, a typical strategy is to employ the divergence theorem to recast this integral in terms of only the gradients of the electronic fields in contrast to Laplacians as in \cref{eqn:fpm,eqn:bgpm}. To this end, we have
\begin{gather}
\begin{multlined}
    \int_{\Omega_p} V_\text{xc}N^{h,p}_IN^{h,p}_J\dr=\\\int_{\Omega_p} \left(f^+N^{h,p}_IN^{h,p}_J+\nabla\left(N^{h,p}_IN^{h,p}_J\right)\cdot\bg^+\right)\dr
\end{multlined}\\
\begin{multlined}
    \int_{\Omega_p} \bB_\text{xc}N^{h,p}_IN^{h,p}_J\dr=\\\int_{\Omega_p} \left(f^-\hat\bmm N^{h,p}_IN^{h,p}_J+\nabla\left(\hat\bmm N^{h,p}_IN^{h,p}_J\right)\cdot\bg^-\right)\dr\label{eqn:divBxc}
\end{multlined}
\end{gather}
While such an approach is satisfactory in the case of spin-unpolarized and collinear-spin calculations and has been employed in prior works~\cite{Motamarri2020DFT-FECalculations, Das2022DFT-FEDiscretization}, it is not viable in the current case of noncollinear framework as this approach requires the evaluation of gradient of the magnetization unit-vector ($\nabla\hat\bmm$), which causes the integral in \cref{eqn:divBxc} to become unbounded when $\bmm=0$. In order to avoid these issues, we resort to \citet{White1994ImplementationCalculations} formalism where the authors prescribe an alternate approach to compute the exchange-correlation potentials for gradient-corrected functionals in the case of plane-wave basis. Extensions of this approach to other basis sets involving atomic-orbital basis, finite-difference techniques can be found in \cite{Bylander1997WhiteAtoms,Balbas2001EvaluationStress,Genovese2008DaubechiesCalculations}. We now propose an extension to the White and Bird approach for evaluating $V_{xc}$ and $\bB_{xc}$ in the case of GGA functionals arising in noncollinear magnetism within the framework of finite-element discretization. To evaluate the integrals in \cref{eqn:FEHloc} using \cref{eqn:cellIntegrals}, we need to compute $V_{xc}$ and $\bB_{xc}$ at the quadrature points and the following discussion provides a prescription to do so.


To this end, we begin with the integral involved in computing exchange-correlation energy ($E_{\text{xc}}$) in \cref{eqn:xcEnergy} and use \cref{eqn:cellIntegrals} to recast it into a discrete form given by
\begin{align}   \widetilde{E}_\text{xc}=\sum_e\sum_{q^\prime}w_{q^\prime}J^{(e)}f_\text{xc}(\rho^\uparrow,\rho^\downarrow,\gamma_0,\gamma_1,\gamma_2)\Biggr\rvert_{\br^{(e)}_{q^\prime}}. \label{eqn:discxcEnergy}
\end{align}
Subsequently, following the prescription in \cite{Bylander1997WhiteAtoms,Balbas2001EvaluationStress,Genovese2008DaubechiesCalculations}, we derive the expressions for the exchange-correlation terms $V_\text{xc}$ and $\bB_\text{xc}$ at quadrature points as
\begin{align}
    V_\text{xc}(\br_q^{(e)})&=\frac{1}{w_qJ^{(e)}}\funcder{\widetilde{E}_\text{xc}}{\rho(\br^{(e)}_q)} \label{eqn:disvxc}\\
    \bB_\text{xc}(\br_q^{(e)})&=\frac{\hat\bmm(\br_q^{(e)})} {w_qJ^{(e)}}\funcder{\widetilde{E}_\text{xc}}{\abs{\bmm}(\br^{(e)}_q)} \label{eqn:disbxc}
\end{align}
We now treat the exchange-correlation energy strictly as a functional of charge density $\rho$ and the magnitude of the magnetization density $\abs{\bmm}$ by defining $\del \rho$ and $\del \abs{\bmm}$ as functionals of $\rho$ and $\abs{\bmm}$ respectively, at the quadrature points. To this end, the exchange-correlation terms $V_\text{xc}$ and $\bB_\text{xc}$ in \cref{eqn:disvxc,eqn:disbxc} can now be recast as
\begin{align}
    V_{\text{xc}}(\br^{(e)}_q)=&f^+(\br^{(e)}_q)+\sum_{q^\prime}\frac{w_{q^\prime}}{w_q}\bg^+(\br^{(e)}_{q^\prime})\cdot\funcder{\nabla\rho(\br^{(e)}_{q^\prime})}{\rho(\br^{(e)}_q)} \label{eqn:vxcwb}\\
    \bB_\text{xc}(\br^{(e)}_q)=&\hat\bmm(\br^{(e)}_q)\Biggl(f^-(\br^{(e)}_q)+\nonumber\\&\sum_{d=x,y,z}\sum_{q^\prime}\frac{w_{q^\prime}}{w_q}\hat m_d(\br^{(e)}_{q^\prime})\bg^-(\br^{(e)}_{q^\prime})\cdot\funcder{\nabla {m_d}(\br^{(e)}_{q^\prime})}{\abs{\bmm}(\br^{(e)}_q)}\Biggr)  \label{eqn:bxcwb}  
\end{align}
where we have used the relation $\nabla\abs{\bmm}=\sum_{d=x,y,z}\hat m_d\nabla m_d$.

We note that the above expressions in \cref{eqn:vxcwb,eqn:bxcwb} require the evaluation of the following functional derivatives
\begin{align}
    \funcder{\nabla\rho(\br^{(e)}_{q^\prime})}{\rho(\br^{(e)}_q)} & &\funcder{\nabla{m_d}(\br^{(e)}_{q^\prime})}{\abs{\bmm}(\br^{(e)}_q)}
\end{align}
where the indices $q$ and $q^\prime$ are over the Gauss-Legendre quadrature points used to evaluate \cref{eqn:discxcEnergy}. To evaluate these functional derivatives, we note that within any given FE-cell the charge density and the magnetization density computed using \cref{eqn:rho,eqn:mag,eqn:FEWfns} lie in the space spanned by 3D tensor-structured polynomials constructed from 1D Lagrange polynomials of order $2n_p$. Consequently, we now define Lagrange polynomial basis $\bar{N}^{(e)}_q(\br) :0\leq q<n_q^3$ constructed over the Gauss-Legendre quadrature points within each FE-cell $\Omega^{(e)}$ with $n_q\geq 2n_p$ so that the total charge density ($\rho$) and the magnetization density ($\bmm = (m_x, m_y, m_z)$) computed using \cref{eqn:rho,eqn:mag,eqn:FEWfns} can be represented exactly using the basis defined by $\bar{N}^{(e)}_q(\br)$ allowing us to express 
\begin{align}
    \rho(\br)&=\sum_q\rho(\br^{(e)}_q)\bar{N}^{(e)}_q(\br) & \forall \br \in \Omega^{(e)}\\
    m_d(\br)&=\sum_q m_d(\br^{(e)}_q)\bar{N}^{(e)}_q(\br)& \forall \br \in \Omega^{(e)}
\end{align}
and correspondingly the gradients 
\begin{align}
    \nabla\rho(\br)&=\sum_q\rho(\br^{(e)}_q)\nabla \bar{N}^{(e)}_q(\br) \\  \nabla m_d(\br)&=\sum_qm_d(\br^{(e)}_q)\nabla \bar{N}^{(e)}_q(\br)
\end{align}
the required functional derivatives can now be evaluated as
\begin{align}
    \funcder{\nabla\rho(\br^{(e)}_{q^\prime})}{\rho(\br^{(e)}_q)} &=\nabla \bar{N}^{(e)}_q(\br^{(e)}_{q^\prime}) \\
    \funcder{\nabla{m_d}(\br^{(e)}_{q^\prime})}{\abs{\bmm}(\br^{(e)}_q)}&=\hat m_d(\br^{(e)}_{q})\nabla \bar{N}^{(e)}_q(\br^{(e)}_{q^\prime})
\end{align}
where we have used the relation ${m_d}=\abs{\bmm}\hat m_d$.
 
\subsubsection{Evaluation of \texorpdfstring{$V_\text{el}$}{the total electrostatic potential}:}
The total electrostatic potential ($V_\text{el}$) is evaluated as the solution of the Poisson problem given by \cref{eqn:poissonEl}. We now define $N^{h,p_\text{el}}_I(\br): 0 \leq I < M^{h,p_\text{el}}$ as the 3D tensor-structured FE polynomial basis constructed from 1D Lagrange polynomials of degree $p_\text{el}$ defined over Gauss Lobatto Legendre (GLL) nodal points~\cite{Brenner2008TheMethods}, generated using the same FE triangulation $\mathcal{T}^h$ used for the solution of \cref{eqn:eigenProbPer}. In this framework, the FE-discretized Poisson problem (\cref{eqn:poissonEl}) reduces to a system of equations given by
\begin{align}
    \bK\bV_\text{el}=\brho\label{eqn:poissonElFE}
\end{align}
Where $\bK$ is the FE discretized Laplacian operator given by
\begin{align}
    K_{IJ}=\int\nabla N_I^{h,p_\text{el}}(\br)\cdot \nabla N_J^{h,p_\text{el}}(\br)\dr
\end{align}
and the $M^{h,p_\text{el}}\times 1$ vector $\bV_\text{el}$ consists of the basis coefficients of the discrete electrostatic potential ($V_\text{el}$) while the $M^{h,p_\text{el}}\times 1$ vector $\brho$ is given by
\begin{align}
    \rho_I=\int \rho(\br)N_I^{h,p_\text{el}}(\br)\dr
\end{align}
The solution to \cref{eqn:poissonElFE} is computed using a conjugate-gradient method.

\subsection{Self-Consistent Field Iteration}
We now discuss the numerical strategies used to solve the discretized nonlinear generalized eigenvalue problem described by \cref{eqn:GHEP,eqn:FEHloc,eqn:FEHnloc,eqn:CMat,eqn:OvlapMap}.
 Towards this goal, we employ the self-consistent field procedure commonly adopted in DFT calculations~\cite{Kresse1996EfficiencySet,Kresse1996EfficientSet,Das2022DFT-FEDiscretization,Motamarri2020DFT-FECalculations,0953-8984-21-39-395502,Giannozzi2017AdvancedESPRESSO} to convert the nonlinear eigenvalue problem to a sequence of linear eigenvalue problems. Each of these linear eigenvalue problems is then solved by making use of a modified Chebyshev filtered subspace iteration (ChFSI) procedure~\cite{Kodali2024AArchitectures}. As the nonlinear eigenvalue problems for each wavevector $\bk$ are mutually independent, we omit the index $\bk$ in the subsequent sections for notational convenience.
\subsubsection{Density Mixing}
The nonlinear eigenvalue problem within the framework of noncollinear magnetism can be formulated as a fixed point iteration problem as
\begin{align}
    F\left[\left(\rho,\bmm\right)\right]=\left(\rho,\bmm\right) \label{eqn:fixedpointmap}
\end{align}
where map $F[\left(\rho_{in},\bmm_{in}\right)]=\left(\rho_{out},\bmm_{out}\right)$ represents the computation of $V_\text{eff}$ and $\bB_\text{xc}$ using $\left(\rho_{in},\bmm_{in}\right)$, solving the FE-discretized eigenvalue problem given by \cref{eqn:GHEP} and computing $\left(\rho_{out},\bmm_{out}\right)$ using \cref{eqn:rho,eqn:mag,eqn:FEWfns}.
In order to accelerate the convergence of the self-consistent field iteration procedure, we employ the Anderson mixing scheme~\cite{Anderson1965IterativeEquations,Eyert1996ASequences}, which computes the input densities for the next iteration as linear combinations of the previous $i$ input densities and residuals.
\begin{align}
    \left(\rho^{i+1}_{in},\bmm^{i+1}_{in}\right)=\sum^i_{j=1}b_{i,j}\left(\rho^{j}_{in},\bmm^{j}_{in}\right)+\alpha b_{i,j}\left(\rho^{j}_{res},\bmm^{j}_{res}\right)
\end{align}
where $\left(\rho^{j}_{res},\bmm^{j}_{res}\right)=F\left[\left(\rho^{j}_{in},\bmm^{j}_{in}\right)\right]-\left(\rho^{j}_{in},\bmm^{j}_{in}\right)$
and $b_{i,j}$ are chosen such that $\sum^i_{j=1}b_{i,j}=1$ and
\begin{align}
    \norm{\sum^i_{j=1}b_{i,j}\left(\rho^{j}_{res},\bmm^{j}_{res}\right)}
\end{align}
is minimized where $\norm{\cdot}$ is the norm induced by the inner product defined as
\begin{align}
    \left<\left(\rho_1,\bmm_1\right),\left(\rho_2,\bmm_2\right)\right>=\frac{1}{2}\left(\int_{\Omega_p}\rho_1\rho_2+\int_{\Omega_p}\bmm_1\cdot\bmm_2\right)\label{eqn:innerPdt}
\end{align}
RPA based preconditioners such as the Kerker~\cite{Kerker1981EfficientCalculations} or the Resta preconditioner~\cite{Resta1977Thomas-FermiSemiconductors} are applied to the total charge density residuals, $\rho_{res}\gets K\rho_{res}$ where $K$ denotes the action of the preconditioner.

\subsubsection{Subspace iteration for linear eigenproblem} In order to solve the generalized Hermitian eigenvalue problem (GHEP) described by \cref{eqn:GHEP}, we employ a Chebyshev filtered subspace iteration (ChFSI) procedure that naturally allows the use of filtered subspace rich in desired eigenvectors of a given SCF iteration to a subsequent iteration progressively improving the convergence towards the self-consistent solution. Furthermore, we note that the ChFSI method is well-suited for modern high-performance computing architectures and has been employed in prior works involving the solution of standard eigenvalue problems arising in real-space DFT calculations~\cite{Motamarri2020DFT-FECalculations,Das2019FastSystem,Zhou2006ParallelAcceleration,Zhou2014Chebyshev-filteredEquation}. To solve GHEP in \cref{eqn:GHEP} using ChFSI, we first seek to amplify the eigenspace of interest by constructing a Chebyshev polynomial filter corresponding to the matrix $\bM^{-1}\bH$ that has the same eigenspace as $\bH\bU = \bM \bU \bLam$. Towards this, we efficiently evaluate $\bM^{-1}$, a $2M^{h,p}\times 2M^{h,p}$ matrix by employing Gauss-Lobatto-Legendre (GLL) quadrature rules coincident with the finite-element nodes of the spectral finite-elements employed in this work, rendering $\bM$ diagonal. Denoting this diagonal matrix as $\bM_D$, we now seek to construct the subspace rich in the desired eigenvectors by first scaling and shifting the matrix $\btH = \bM_D^{-1}\bH$ so that the unwanted spectrum ($\left[\lambda_T,\lambda_{max}\right]$) is mapped to $\left[-1,1\right]$ where $\lambda_T$ and $\lambda_{max}$ denote the upper bounds of wanted and unwanted spectrum respectively. Subsequently, the filtered vectors are evaluated using the three-term recurrence relation of the Chebyshev polynomials as
\begin{align}
    &\bX_k=T_k(\btH)\bX \;\; \forall k=0,1,2,\dots,s \nonumber\\
    &\text{where}\;\; T_{k+1}(\btH) = 2\btH T_k(\btH) - T_{k-1}(\btH) \label{eqn:chebfilt}
\end{align}
where $\bX$ is the initial guess of the eigenvectors and $T_k$ is the scaled Chebyshev polynomial of degree $k$, and $s$ denotes the choice of the Chebyshev polynomial degree used for filtering the vectors in a given SCF iteration. The above recurrence relation involving the scaled Hamiltonian exploits the exponential growth of Chebyshev polynomials outside of $[-1,1]$ to amplify the desired spectrum $\left[\lambda_{min},\lambda_T\right)$ where $\lambda_{min}$ denotes the lower bound of the wanted spectrum. Values of $\lambda_{min}$ and $\lambda_{max}$ are estimated based on the Lanczos method with a generalized variant~\cite{vanderVorst1982AScheme} while a good approximation to the value of $\lambda_T$ is estimated based on the highest generalized Rayleigh quotient of $\btH$ of the previous SCF iteration. 

In the conventional Chebyshev filtering algorithm as described above, the predominant computational cost is the evaluation of Hamiltonian times vector products, $\bH\bX_k$. To this end, we reformulate the filtering step\footnote{Manuscript under preparation} by substituting these products with Hamiltonian-times residual products, and as we approach convergence, these eigenproblem residuals progressively become smaller. Consequently, it becomes viable to compute these Hamiltonian times residual products in lower precision without compromising accuracy, thus improving the computational efficiency~\cite{Kodali2024AArchitectures}.

Once we obtain the filtered subspace $\bX_s=T_s(\btH)\bX$, we use the Rayleigh-Ritz projection step to obtain the desired eigenvectors and eigenvalues. In light of this, we solve the subspace-projected eigenvalue problem defined by
\begin{align}
    \bX_s^\dagger\bH\bX_s\bQ_s=\bX_s^\dagger\bM\bX_s\bQ_s\bLam_s
\end{align}
where $\bQ_s$ and $\bLam_s$ are the eigenvectors and eigenvalues of the above $T\times T$ generalized eigenvalue problem. The new estimates for the eigenvectors and eigenvalues can be obtained as
\begin{align}
    \bU&=\bX_s\bQ_s& \bLam&=\bLam_s
\end{align}
This process of Chebyshev filtering, followed by a Rayleigh-Ritz step, is repeated till a desired residual tolerance is reached for a given SCF iteration.

\section{\label{sec:implem} Numerical implementation strategies}


In this section, we discuss the computationally efficient approaches underlying our numerical implementation that accelerate the finite-element (FE) based solution procedure described in \cref{sec:computemethod}. In particular, we describe the two-grid strategy employed to accelerate the SCF procedure and further highlight the key numerical aspects involved in evaluating the action of FE discretized Hamiltonian on the trial subspace of complex spinors, the computationally intensive step of the linear eigensolver within the framework of noncollinear magnetism.

\subsection{Two-grid strategy}
At the beginning of the self-consistent field (SCF) iteration where the electronic fields $\rho$ and $\bmm$ are far away from the self-consistent solution, one does not need to evaluate $ F\left[\left(\rho,\bmm\right)\right]$ in \cref{eqn:fixedpointmap} accurately, and leveraging this, we solve the linear eigenvalue problem approximately using a lower FE interpolating polynomial order $p-1$ in the initial SCF iterations instead of using $p$. We refer to this approach as the ``Two-grid strategy" and here, the discretized 2-component complex spinors are expressed as
\begin{align}
    \bu_{n\bk}^{h,p-1}\left(\br\right)&=\sum_{I}\bu_{n\bk}^IN^{h,p-1}_I(\br)\label{eqn:FEWfnsCMesh}
\end{align}
where $N^{h,p-1}_I(\br): 0 \leq I < M^{h,p-1}$ are the 3D tensor-structured FE polynomial basis constructed from 1D Lagrange polynomials of degree $p-1$ defined over Gauss Lobatto Legendre (GLL) nodal points~\cite{Brenner2008TheMethods}, generated using the nodes of the same FE triangulation $\mathcal{T}^h$ defined in \cref{sec:FEDiscretization}. Consequently the FE discretized eigenvalue problem defined by \cref{eqn:GHEP} and the FE discretized matrices defined by \cref{eqn:FEHloc,eqn:FEHnloc,eqn:CMat,eqn:OvlapMap} are also computed using  $N^{h,p-1}_I(\br)$ instead of $N^{h,p}_I(\br)$ resulting in an eigenvalue problem of reduced dimension ($ M^{h,p-1}< M^{h,p}$) that is computationally cheaper to solve. We perform the self-consistent iteration using this reduced-order discretization until a chosen tolerance of charge and magnetization density residuals (in the norm induced by the inner product defined by \cref{eqn:innerPdt}) is achieved. In all the numerical calculations reported in this work, this tolerance is typically chosen higher than the value chosen for the convergence of the complete self-consistent iteration.

\subsection{Action of FE discretized Hamiltonian}

We note that the action of the FE discretized Hamiltonian on a trial subspace encountered in the ChFSI procedure described previously involves the computationally intensive step of evaluating the matrix-multivector product $\bH\bY$ that entails the multiplication of a sparse-matrix ($\bH$) with a dense-matrix ($\bY$). It has been noted previously that a direct sparse-matrix times dense-matrix (spMM) is computationally expensive than other methods available in the FE literature~\cite{Cantwell2011FromElements,Carey1988Element-by-elementComputations,Hughes1987Large-scaleGradients,Das2023Large-ScaleAlloys,Motamarri2020DFT-FECalculations,Das2022DFT-FEDiscretization,Das2019FastSystem}. We adopt one such method by extending the strategy to the FE discretized matrices arising in the noncollinear and the SOC framework described here, where the spMM operation is recast into multiple smaller dense-matrix times dense-matrix products that have high arithmetic intensity. Key ideas are described below.
\vspace{0.2in}
\subsubsection{Cell-level Hamiltonian matrices:}
Towards the goal of recasting to small dense matrix-matrix products, we now define dense FE cell-level matrix ${\bH^\text{loc}}^{(e)}$ of size $2n_p^3\times 2n_p^3$ corresponding to the local part of the FE discretized Hamiltonian $\bH^\text{loc}$ defined in \cref{eqn:FEHloc} in the following manner
\begin{align}
    \bH^{\text{loc}}=\sum_{e}{\bRcal^{(e)}}^T{\bH^\text{loc}}^{(e)}\bRcal^{(e)}
\end{align}
where $\bRcal^{(e)}$ is a $2M^{h,p}\times 2n_p^3$ Boolean sparse matrix which extracts the  basis coefficients, $\bu_{n\bk}^I$ in \cref{eqn:FEWfns}, corresponding to an FE-cell $\Omega^{(e)}$, i.e. $I\in\Omega^{(e)}$ and is commonly referred to as the restriction matrix. 

In this framework the cell-level non-local projector matrices ${\bPt^a}^{(e)}$ of size $2n_p^3 \times 2n_{\text{pj}}^a$ corresponding to the FE discretized projector matrices defined by \cref{eqn:CMat} can be written as
\begin{align}
    \begin{bmatrix}\bPt^a &0\\0& \bPt^a\end{bmatrix}=\sum_e{\bRcal^{(e)}}^T{\bPt^a}^{(e)}
\end{align}
Furthermore, we define the $2n_{\text{pj}}^a\times 2n_{\text{pj}}^a$ matrix $\bDel^{\gamma_a}$ corresponding to the non-local pseudopotential coefficients present in the non-local part of the Hamiltonian (\cref{eqn:FEHnloc}) as
\begin{align}
    \Delta^{\gamma_a}_{2\rchi+\alpha,2\rchi^\prime+\beta}=D^{\gamma_a,\rchi,\rchi^\prime}_{\alpha\beta}
\end{align}

The cell-level local Hamiltonian matrices (${\bH^\text{loc}}^{(e)}$) and the non-local projector matrices ${\bPt^a}^{(e)}$  are computed as
\begin{multline}    
    \text{H}_{2I+\alpha,2J+\beta}^{{\text{loc}}^{(e)}}=\int_{\Omega^{(e)}}\Biggl(\biggl[\frac{1}{2}\nabla N^{h,p}_I(\br)\cdot\nabla N^{h,p}_J(\br)\\+ V_\text{eff}(\br)N^{h,p}_I(\br)N^{h,p}_J(\br)\biggr]\delta_{\alpha\beta}\\+\sum_{d=x,y,z}\biggl[B^d_\text{xc}(\br)N^{h,p}_I(\br)N^{h,p}_J(\br)\biggr]\sigma^d_{\alpha\beta}\Biggr)\dr\label{eqn:FEHlocEl}
\end{multline}
\begin{align}    
{\text{P}^a}^{(e)}_{2I+\alpha,2\rchi+\beta}=\delta_{\alpha\beta}\int_{\Omega^{(e)}}p^a_{\rchi}(\br)N^{h,p}_I(\br)\dr\label{eqn:FEHnlocEl}
\end{align}
In order to evaluate these integrals we define the Lagrange polynomial basis functions, $\widetilde{N}_I$, defined on 3D tensor structured Gauss-Legendre-Lobatto quadrature points, $\widetilde{\br}_q$ in the reference cell $\widetilde{\Omega}=\left[-1,1\right]^3$. In order to elucidate the methodology followed for the  evaluation of the integrals in \cref{eqn:FEHlocEl,eqn:FEHnlocEl} we consider the term $\int V_\text{eff}N^{h,p}_IN^{h,p}_J$, in terms of the basis functions in the reference cell this integral can be written as
\begin{align}
    \int_{\Omega^{(e)}}V_\text{eff}(\br)&N^{h,p}_I(\br)N^{h,p}_J(\br)\dr\\=&\int_{\widetilde{\Omega}}V_\text{eff}(\br(\widetilde{\br}))\widetilde{N}^{h,p}_I(\widetilde{\br})\widetilde{N}^{h,p}_J(\widetilde{\br})J^{(e)}d\widetilde{\br}\\=&\sum_{q}V_\text{eff}(\br_q)\widetilde{N}^{h,p}_I(\widetilde{\br}_q)\widetilde{N}^{h,p}_J(\widetilde{\br}_q)J^{(e)}w_q
\end{align}
Defining $\widetilde{\bN}$ to be the $n_p^3\times n_q^3$ matrix whose elements are given by $\widetilde{\text{N}}_{Iq}=\widetilde{N}_I(\br_q)$ and $\bV^\text{eff}$ to be the $n_q^3\times n_q^3$ diagonal matrix whose elements are given by $\text{V}^\text{eff}_{qq^\prime}=\delta_{qq^\prime}w_qJ^{(e)}V_\text{eff}(\br_q)$, the integral can now be written as
\begin{align}
    \int_{\Omega^{(e)}}V_\text{eff}(\br)N^{h,p}_I(\br)N^{h,p}_J(\br)\dr&=\text{V}_{IJ}
\end{align}
where the matrix $\bV$  is given by
\begin{align}  \bV&=\left(\widetilde{\bN}\circ\bV^\text{eff}\right)\widetilde{\bN}^T
\end{align}
where $\circ$ represents the Hadamard product of two matrices. The matrix $\bV$ can now be efficiently evaluated using standard level-3 BLAS functions and their strided/batched variants. The other integrals in \cref{eqn:FEHlocEl,eqn:FEHnlocEl} are also evaluated using a similar methodology.
\vspace{0.2in}
\subsubsection{FE discretized matrix multivector product}
The matrix multivector product ${\bY} = \bH {\bX}$ required in \cref{eqn:chebfilt} can then be evaluated using these FE-cell level dense matrices and the FE-cell level multivectors~\cite{Das2022DFT-FEDiscretization,Motamarri2020DFT-FECalculations,Carey1988Element-by-elementComputations}. This strategy comprises of the following steps~:
\begin{itemize}
    \item [1.] \underline{Precompute} the FE-cell level operator matrices ${\bH^\text{loc}}^{(e)}$ and ${\bPt^a}^{(e)}$.\label{stp:setupele} 
    \item [2.] \underline{Extraction} of the FE-cell level multivectors ${\bY}^{(e)}$ using the restriction matrix,
    i.e., ${\bX}^{(e)}=\bRcal^{(e)}\bX \\ \forall e=1,2,\dots,E$.
    \item [3.] \underline{FE-cell level evaluation} of the matrix multivector products 
    \begin{align}
        {\bY}^{\left(e\right)}={\bH^\text{loc}}^{(e)}{\bX}^{(e)}+\sum_a{\bPt^a}^{(e)}\bDel^{\gamma_a}\sum_{e^\prime}{{\bPt^a}^{(e^\prime)}}^\dagger{\bX}^{(e^\prime)}
    \end{align}
    This operation is done using BLAS routines~\cite{2002AnBLAS} for dense matrix-matrix multiplication.\label{stp:elecompute}
    \item [4.] \underline{Assembly} of the global multivector $\bY$ using the restriction matrix, i.e., ${\bY} = \sum_{e}^{E}{\bRcal^{(e)}}^T\bY^{(e)}$
\end{itemize}
\section{\label{sec:benchmarks} Results and Discussion}
We now present comprehensive studies demonstrating the accuracy, performance and parallel scalability of the proposed computational approach (\ncsocfe{}) on model benchmark systems involving periodic, semi-periodic and non-periodic boundary conditions.  Accuracy benchmarks involve comparisons of ground-state energies, magnetic anisotropy energies, volume integral of magnetization densities, vertical ionization potentials, spin textures, band-structures with plane-wave-based DFT calculations using Quantum espresso (\qe{})~\cite{Giannozzi2009QUANTUMMaterials.,Giannozzi2017AdvancedESPRESSO,Carnimeo2023scpQuantumExascale} for various representative examples considered in this work. Furthermore, on CPU architectures, performance comparisons of \ncsocfe{}  with respect to plane-wave DFT calculations involving noncollinear magnetism and spin-orbit coupling have been carried out for system sizes ranging from $\sim$ 3000 to 15000 electrons. To this end, we employ two metrics to compare the performance: (i) computational cost per SCF iteration \footnote{$\eta$ is obtained by multiplying the minimum number of compute nodes required to fit a given problem with the average wall-time per SCF iteration} ($\eta$) in node-hrs, and (ii) minimum wall time per SCF iteration \footnote{$\tau^{\text{min}}$ is obtained by computing the average wall time per SCF by increasing the number of compute nodes till the time does not change significantly or starts increasing} ($\tau^{\text{min}}$) in secs. Additionally, we present performance benchmarks of our method on GPU architectures for these large-scale systems. Finally, we showcase the parallel scalability of \ncsocfe{} on multi-node CPU and GPU architectures on representative periodic and semi-periodic material systems. All  simulations involved in the accuracy benchmarking are performed on KNL CPU nodes on Nurion\footnote{Nurion is one of South Korea's fastest supercomputers stationed at KIST comprising of 8305  Intel Xeon KNL based CPU nodes (564,740 Cores) where each node consists of 68 cores (Intel Xeon Phi 7250 processor), 96 GB memory and Fat-tree topology based high-performance interconnect between all the nodes for fast MPI communication.}, and performance benchmarking studies were performed both on
CPU nodes of Nurion and GPU nodes of Frontier\footnote{Frontier is the world's first exascale supercomputer stationed at ORNL comprising of 9408 AMD compute nodes with node containing 8 GPUs per node each having 64 GB of high-bandwidth memory}. Note that we do not show GPU benchmarks of \qe{} on Frontier as it does not yet fully support AMD GPUs~\cite{Ruffino2024QuantumOpenMP}.

Unless otherwise specified, all the DFT calculations reported in this work utilize the PBE functional~\cite{PhysRevLett.77.3865} for the exchange-correlation and fully relativistic optimized norm-conserving pseudopotentials (ONCV)~\cite{Hamann2013OptimizedPseudopotentials} from the Pseudo-Dojo database~\cite{vanSetten2018TheTable}. All calculations were performed using Fermi-Dirac smearing with a smearing temperature of 500 K. The initial guess for magnetization density is chosen to be similar for both Quantum-Espresso and DFT-FE for all the calculations reported here. In the accuracy validation studies discussed in this work using \ncsocfe{}, we employ refined finite element (FE) meshes with FE interpolating polynomial ($p$) of degree 6. These meshes are constructed such that the discretization error in ground-state energy is less than $\order{(10^{-5})}$ Ha/atom. Additionally,  FE interpolating polynomial degree for electrostatics $p_{\text{el}}$ and the quadrature integration rules are chosen such that the energy variation with respect to these parameters is an order of magnitude lower than this discretization error. Additionally, for mixing of the electron charge density $\rho$ and magnetization density $\bmm$, we employ the $n$-stage Anderson mixing scheme~\cite{Anderson1965IterativeEquations} as discussed in the \cref{sec:computemethod}. In the case of plane wave calculations for accuracy validation studies using \qe{}, the cutoff energy for wavefunctions $\texttt{ecutwfc}$ is chosen so that the discretization error in ground-state energy is less than $\order{(10^{-5})}$ Ha/atom while simultaneously ensuring that the change in energy with respect to the cutoff energy for the electron charge density $\texttt{ecutrho}$ is an order of magnitude lower. Further, the default mixing scheme is used in the case of \qe{}. For metallic systems we use the Kerker preconditioner in both \qe{} and \ncsocfe{}. The structures of all the systems considered for accuracy benchmarks can be found in the supplementary material.


\subsection{Non-periodic/Semi-periodic systems}
In this subsection, we examine the case of fully non-periodic and semi-periodic systems for accuracy and performance benchmarking. In particular, we benchmark the accuracy of \ncsocfe{}  with \qe{} by comparing the relaxed ground state energies of isolated systems and a semi-periodic system involving TMD bilayer. Furthermore, for some of these materials systems, we compute the vertical ionization potential and the volume integral of the magnetization density and compare the values with that obtained from \qe{}. Following the accuracy validation study, we evaluated the performance of our implementation by comparing the computational cost in node hours and the minimum wall time of \ncsocfe{}  with \qe{}, a widely used DFT code that uses the plane-wave basis. In this performance benchmarking study, we consider semi-periodic systems involving WTe$_2$ for various twist angles ranging from 180 atoms (3600 electrons) to 1032 atoms (20640 electrons). In \ncsocfe{}, we apply homogenous Dirichlet boundary conditions in the non-periodic directions for charge-neutral systems when solving for the electrostatic potential using \cref{eqn:poissonElFE}, and a suitable vacuum is used till the electronic fields decay to 0 in these directions. In the plane-wave code \qe{}, periodic boundary conditions (PBCs) are only admissible; hence, PBCs are employed using a suitable vacuum to minimize image-image interactions in non-periodic directions.

\subsubsection{Accuracy benchmarking}
In all the accuracy benchmarking studies of \ncsocfe{} against \qe{} reported here, we employ a kinetic energy cutoff (\texttt{ecutwfc}) of 75Ha in \qe{} and a polynomial order of 6 with a finite-element mesh size of 0.8 Bohr in our method.

\paragraph{Energetics:} 
\Cref{tab:intEnergynonper} shows the comparison of total internal energies with \qe{}, for a few representative gas-phase molecules from the GW-SOC81 benchmark set~\cite{Scherpelz2016ImplementationSolids} involving heavy atoms (AgBr, AsI$_{3}$, (C$_5$H$_5$)$_2$ Ru)), 3-atom cluster of Chromium Cr$_3$ that exhibits a noncollinear magnetic state and bilayer WTe$_2$, a semi-periodic system that displays strong spin-orbit effects. In \qe, for all the isolated systems simulated in this study, the Gamma point is used to sample the Brillouin zone. While simulating the semi-periodic system WTe$_2$ in \qe, a shifted $4 \times 4 \times 1$ Monkhorst-Pack k-point grid is used for sampling the Brillouin zone~\cite{Monkhorst1976SpecialIntegrations} and periodic boundary conditions is applied in all 3-directions. In the case of \ncsocfe{}, we employ periodic boundary conditions only in two directions spanning the plane of the bilayer, employing a shifted $4 \times 4 $ Monkhorst-Pack k-point grid to sample the Brillouin zone.  
\begin{table}[!h]
    \makegapedcells
    \setlength\tabcolsep{2pt}
    \begin{tabularx}{\linewidth}{|Y|Y|Y|Y|}
    \Xhline{2pt}
    System & \qe{} (Ha) & \ncsocfe{} (Ha) & Error (Ha/atom)\\
    \Xhline{2pt}     
    AgBr   & -170.704020&-170.703937 &4.1E-05\\
    AsI$_3$   & -144.897928&-144.897902 &6.6E-06\\
    (C$_5$H$_5$)$_2$Ru   & -167.114509&-167.114752 &1.2E-05\\
    Cr$_3$   &-266.639352 &-266.639406 &1.8E-05\\
    WTe$_2$   & -1143.255383&-1143.254923 &3.8E-05\\
    \Xhline{2pt}
    \end{tabularx}
    \caption{\label{tab:intEnergynonper}Accuracy benchmarks for total internal energy.}
\end{table}
\begin{table}[!b]
    \makegapedcells
    \setlength\tabcolsep{2pt}
    \begin{tabularx}{\linewidth}{|Y|Y|Y|Y|}
    \Xhline{2pt}
    System & \qe (Bohr Magneton) & \ncsocfe{} (Bohr Magneton) & Error (Bohr Magneton)\\
    \Xhline{2pt}
    Cr$_3$   & 14.5367&14.5358 &8.9E-04\\
    \Xhline{2pt}
    \end{tabularx}
    \caption{\label{tab:absMagCr3}Accuracy benchmarks for volume integral of the magnitude of magnetization density.}
\end{table}

From \cref{tab:intEnergynonper}, we observe an excellent match between \ncsocfe{} and \qe{}, and the errors between the approaches are well within chemical accuracy, validating the accuracy of our implementation. Furthermore, we also compare the volume integral of the magnitude of magnetization density for the 3-atom cluster of Cr$_3$ in \cref{tab:absMagCr3} and find a close correspondence between the two approaches compared here.

\paragraph{Verical Ionization potentials:} We now compare the vertical ionization potentials of a few gas-phase molecules from the GW-SOC81 benchmark set~\cite{Scherpelz2016ImplementationSolids}. The vertical ionization potentials are computed using the $\Delta$-SCF method
\begin{align}
    \text{VIP}=E(N_e-1)-E(N_e)
\end{align}
where $N_e$ is the number of valence electrons in the neutral molecule. To this end, we need to perform DFT calculations for charged molecules in both the approaches compared here. In the case of ~\ncsocfe{}, we accomplish this by imposing multipole boundary conditions (up to the quadrupole term) on the total electrostatic potential while solving \cref{eqn:poissonEl} in a sufficiently large domain. In \qe{}, we utilize the Makov-Payne correction~\cite{Makov1995PeriodicCalculations} for isolated systems with sufficiently large supercells to avoid image interactions. Note that the treatment of the charged system is not equivalent to that of \qe{} in our framework, and as such, we do not expect an exact match with it.
\begin{table}[!h]
    \makegapedcells
    \setlength\tabcolsep{2pt}
    \begin{tabularx}{\linewidth}{|Y|Y|Y|Y|}
    \Xhline{2pt}
    System & \qe (eV) & \ncsocfe{} (eV) & Experiment (eV)\cite{Scherpelz2016ImplementationSolids,Linstrom2024NIST69}\\
    \Xhline{2pt}
    AgBr   & 9.53&9.44 &9.59\\
    AsI$_3$   & 8.58&8.62 &9.00\\
    (C$_5$H$_5$)$_2$Ru   &7.23 &7.26 &7.45\\
    \Xhline{2pt}
    \end{tabularx}
    \caption{\label{tab:vip}Accuracy benchmarks for vertical ionization potentials.}
\end{table}

\begin{figure}[!hbt]
    \includegraphics[width=\linewidth]{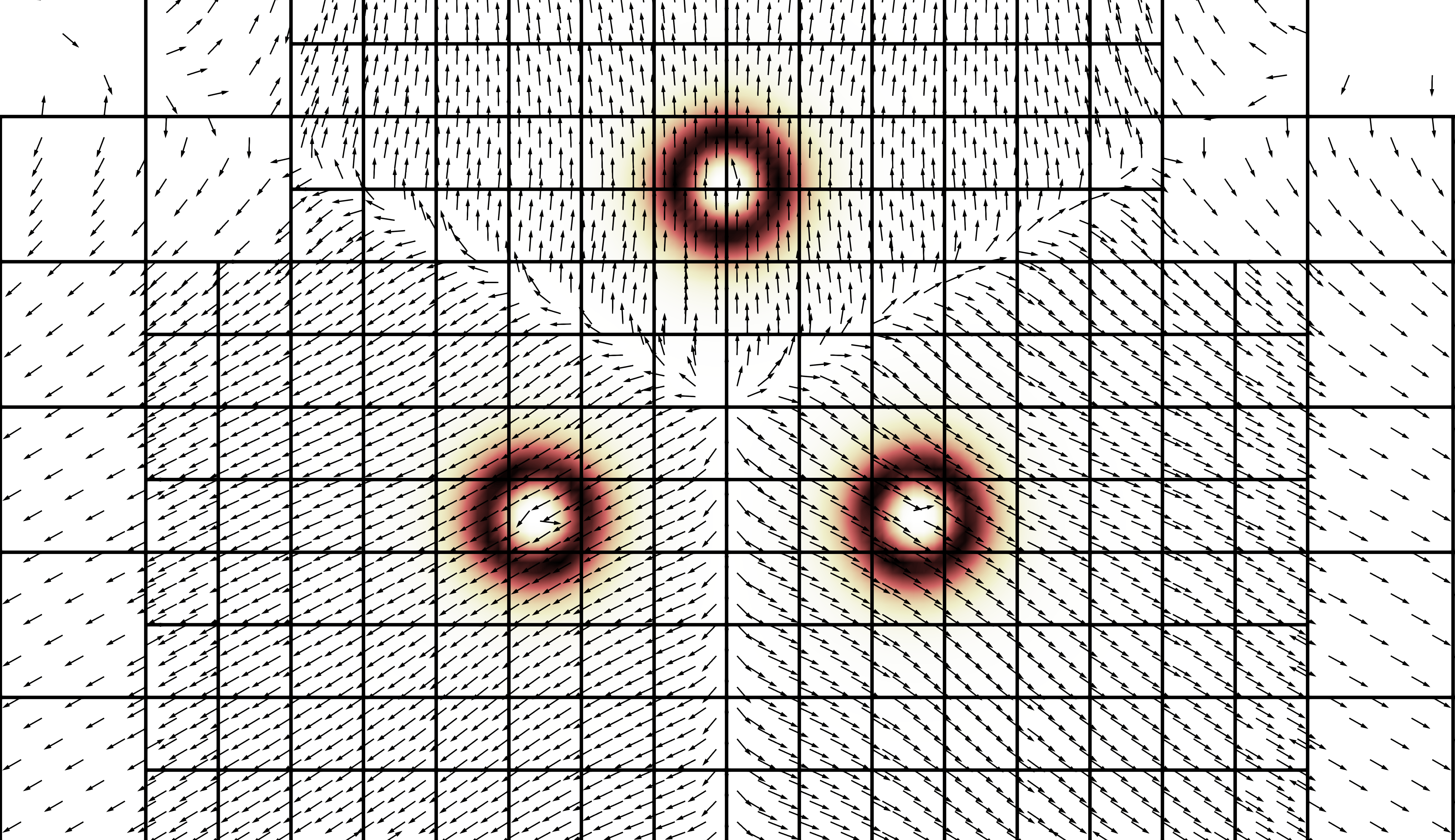}
    \caption{Spin-textures for Cr$_3$}\label{fig:spinTex}
\end{figure}

\paragraph{Spin textures:} We also illustrate the spin texture of the Cr$_3$ system, which is known to exhibit a noncollinear magnetic Neel state due to geometric frustration. To this end, we consider a fully relaxed 3-atom cluster of Chromium (Cr$_3$); resulting in a Cr-Cr bond length of 4.86~Bohr and plot the spin-texture of the Neel state of Cr$_3$ in \cref{fig:spinTex}
We note that the spin texture obtained from our implementation is consistent with existing studies~\cite{Naveh2009Real-spaceTheory,Pu2023NoncollinearTheory}.
\begin{table*}[t]
    \makegapedcells
    \setlength\tabcolsep{2pt}
    \begin{tabularx}{\linewidth}{|Y|Y|Y|Y|Y|Y|Y|Y|}
        \Xhline{2pt}
       System  & Electrons& Twist angle&\qe{}  \# of basis fns.& \qe{} Nurion CPU Node-hrs ($\eta_{\text{c}}$) & \ncsocfe{}  \newline \# of  basis fns. & \ncsocfe{}\newline Nurion CPU Node-hrs ($\eta_{\text{c}}$) & \ncsocfe{}\newline Frontier GPU Node-hrs ($\eta_{\text{g}}$)\\
        \Xhline{2pt}
         W$_{60}$Te$_{120}$ & 3600 & 30$^\circ$ & 298615 & 0.9 & 3215206 & 2.9 & 0.04 \\
         \hline
         W$_{116}$Te$_{232}$ & 6960 & 49$^\circ$& 577425 &8.2  & 6509526 & 14.0 & 0.2\\
         \hline
         W$_{168}$Te$_{336}$ &10080 & 26$^\circ$& 836149 & 37.8 & 8978220 & 27.8 & 0.4 \\
         \hline
          W$_{228}$Te$_{456}$ &13680 & 15.5$^\circ$& 1134763 & 102.6 & 11971630 & 56.7 & 0.9 \\
         \hline
          W$_{344}$Te$_{688}$ &20640 &12.3$^\circ$ & 1711991 & --\footnote{We were unable to perform this computation due to memory limitations on KNL architectures} & 17930753 & 223.5 & 3.0 \\
          \Xhline{2pt}
    \end{tabularx}
    \caption{Computational cost ($\eta$) comparison between \ncsocfe{} and \qe{} in node-hrs (discretization error $\sim 10^{-4}$~Ha/atom). $\eta$ is computed from the minimum number of nodes required to fit a given material system and the average wall time taken per SCF iteration. \textbf{Case Study}: Twisted  bilayers of WTe$_2$ with varying twist angles}
    \label{tab:computecostwte2}
\end{table*}

\subsubsection{Performance studies}
We now discuss the performance of \ncsocfe{} with \qe{} by comparing the average CPU time per SCF iteration in terms of computational node-hrs ($\eta_{\text{c}}$) with increasing system sizes of a semi-periodic system solved to a similar level of accuracy. In particular, we consider the twisted bilayer WTe$_2$ for various twist angles. Incommensurate twisted bilayer WTe$_2$ requires simulation cells with a large number of atoms, and we consider twist angles ranging from 30 degrees to 12 degrees. To this end, semi-periodic pseudopotential DFT calculations involving noncollinear magnetism and spin-orbit coupling terms are conducted on W$_{60}$Te$_{120}$ (3600 electrons), W$_{116}$Te$_{232}$ (6960 electrons), W$_{168}$Te$_{336}$ (10080 electrons), W$_{228}$Te$_{456}$ (13680 electrons) and W$_{344}$Te$_{688}$ (20640 electrons). The structures were generated using the methodology described by \citet{He2021GiantWTe2}. \Cref{tab:computecostwte2} summarizes the twist angles and the configurations considered. 


We employ the PBE functional~\cite{PhysRevLett.77.3865} for the exchange-correlation and fully-relativistic optimized norm-conserving pseudopotentials (ONCV)~\cite{Hamann2013OptimizedPseudopotentials} in the SG15 database~\cite{Scherpelz2016ImplementationSolids}. In \ncsocfe{}, we use the degree of FE interpolating polynomial $p$ to be 7 and a FE mesh size of 1.5~Bohr while we employ a kinetic energy cutoff (\texttt{ecutwfc}) of 55~Ry in \qe{}. These discretization parameters are chosen so that the discretization error in the ground-state energies obtained in both \ncsocfe{} and \qe{} is $\sim \order{(10^{-4})}$~Ha/atom. Additionally, we perform a non-magnetic calculation with SOC in both \ncsocfe{}  and \qe{} employing a vacuum of around 14~Bohr and 10~Bohr, respectively above and below the twisted bilayer system, ensuring that ground-state energies are converged up to $\order{(10^{-5})}$~Ha/atom with vacuum size. Periodic boundary conditions are applied in the two lattice vector directions spanning the plane of the bilayer, and homogeneous Dirichlet boundary conditions are applied on the electrostatic potential in the direction normal to the bilayer in \ncsocfe{}.  In the case of \qe{} periodic boundary conditions are employed in all three directions and Gamma point sampling of the Brillouin zone is used.

Table~\eqref{tab:computecostwte2} reports the average CPU time per SCF iteration in terms of computational node-hrs ($\eta_{\text{c}}$) and the number of basis functions in \ncsocfe{}  and \qe{} for various sizes of twisted bilayers of WTe$_2$ considered here. Further, we also report the average time per SCF iteration in terms of GPU node-hrs ($\eta_{\text{g}}$) in this table. From this table, we find that for system sizes $\sim 10,000$ electrons and above, \ncsocfe{}  becomes more efficient than \qe{} and gains increase with system size on CPUs. This increase in computational gains is attributed to the need for using more processors to satisfy the peak memory requirement, where the efficient parallel scalability of \ncsocfe{} provides the necessary advantage. Additionally, we estimate the computational complexity in the regime of $\sim3000-20000$ electrons for \ncsocfe{} from table~\eqref{tab:computecostwte2} and is found to be of $\order{(N_e^{2.4})}$ while for \qe{} we observe it to be of $\order{(N_e^{3.6})}$. Hence, we expect to see further gains of \ncsocfe{} over \qe{} with increasing system size on CPUs. We further note from this table that we obtain significant computational efficiencies on GPUs using the proposed computational methodologies. In particular, we find  $\sim$ 60$\times-70\times$ Nurion CPU node-hr to Frontier GPU node-hr speedups, underscoring the importance of the numerical strategies developed in this work amenable for efficient implementation on GPUs as well.

\subsubsection{Scalability}
\begin{figure}[!b]
    \includegraphics[width=\linewidth]{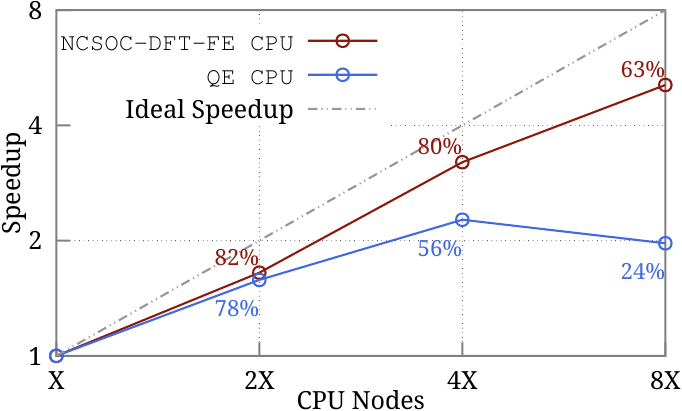}
    \caption{Relative speed up of wall-time per SCF iteration for the 49$^\circ$ twist angle bilayer WTe$_2$ for varying number of KNL CPU nodes on Nurion. \ncsocfe{}: The value of X is 100, and the total number of DoFs is 6509526. The number of band groups at X, 2X, 4X, and 8X is 1, 1, 2, and 4, respectively. \qe{}: The value of X is 20, and the number of plane waves is 577425. Number of band groups at X, 2X, 4X, and 8X is 1, 1, 2, and 2 respectively}\label{fig:wte2ScalCPU}
\end{figure}

\begin{figure}[!t]
    \includegraphics[width=\linewidth]{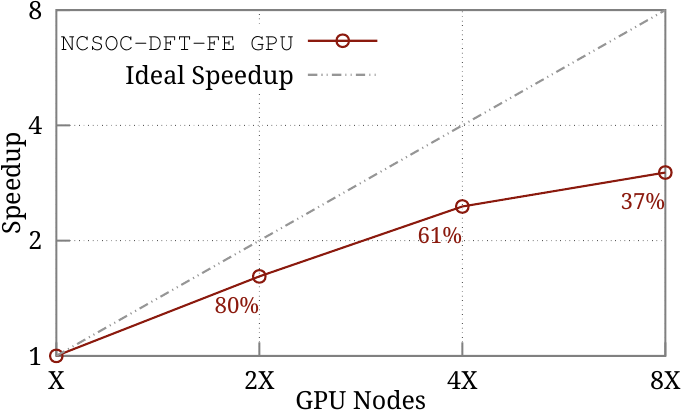}
    \caption{Relative speed up of wall-time per SCF iteration for the 49$^\circ$ twist angle bilayer WTe$_2$ for varying number of GPU nodes on Frontier. \ncsocfe{}: The value of X is 8, and the total number of DoFs is 6509526.}\label{fig:wte2ScalGPU}
\end{figure}

We now assess the parallel scalability (strong scaling) of our numerical implementation involving noncollinear magnetism and spin-orbit coupling (\ncsocfe{}) on both multi-node CPUs and GPUs. We choose the W$_{116}$Te$_{232}$ system containing around 6.5 million degrees of freedom (number of finite-element basis functions) as our benchmark system and present the relative speed ups with respect to the minimum number of both CPU and GPU nodes the problem could fit on. \Cref{fig:wte2ScalCPU} compares the scaling behavior of \ncsocfe{} with \qe{} on KNL CPU nodes and \cref{fig:wte2ScalGPU} demonstrates the scalability of \ncsocfe{} on Frontier AMD GPUs. In both \ncsocfe{} and \qe{}, the discretization parameters are chosen such that the discretization error in the ground-state energy is $\sim \order{(10^{-4})}$ Ha/atom as in previous studies. As evident from the figure, we obtain a scaling efficiency of $\sim$63 \% even at 800 KNL CPU nodes and obtain an efficiency of $\sim$40 \% on 64 GPU nodes (512 GPUs). In the case of \qe{}, we find that the relative speedups obtained by increasing the number of nodes are of a limited range and flatten off early with relative speedups dropping substantially beyond a certain number of CPU nodes compared to \ncsocfe{}. 
Using these results, \cref{tab:computecostwte2minwall} compares the minimum wall time per SCF obtained using \ncsocfe{} with \qe{} for the twisted bilayer W$_{116}$Te$_{232}$ considered in this study. To this end, we find that our method provides nearly $8 \times$ computational gains over \qe{} in terms of minimum wall time on multi-node CPUs and more than $30 \times$  on multi-node GPUs, and these gains will increase substantially with system sizes.
\begin{table}[!h]
    \makegapedcells
    \setlength\tabcolsep{2pt}
    \begin{tabularx}{\linewidth}{|Y|Y|Y|}
        \Xhline{2pt}
        Method & \# of nodes & Minimum Walltime per SCF ($\tau^\text{min}$)\\
        \Xhline{2pt}
       \qe{} Nurion CPUs& 80& 740\\
       \hline
       \ncsocfe{} Nurion CPUs & 800&99\\
       \hline
       \ncsocfe{} Frontier GPUs & 64&26\\
        \Xhline{2pt}
    \end{tabularx}
    \caption{Computational cost ($\tau^\text{min}$) comparison between \ncsocfe{} and \qe{} in terms of minimum walltime (discretization error $\sim 10^{-4}$~Ha/atom). \textbf{Case Study}: 49$^\circ$ twist angle bilayer WTe$_2$}
    \label{tab:computecostwte2minwall}
\end{table}

\subsection{Periodic systems}
We now discuss the case of fully periodic systems for accuracy and performance benchmarking. In particular, we benchmark the accuracy of \ncsocfe{}  with \qe{} by comparing the ground-state energies, volume integral of magnetization density, magnetic anisotropy energies and the bandstructure of representative periodic systems. We further assess the performance of our implementation by comparing the computational cost of \ncsocfe{} with \qe{} by considering periodic systems involving MnSi supercells of various sizes ranging from 288 atoms (2736 electrons) to 1568 atoms (14896 electrons). In \ncsocfe{}, we apply periodic boundary conditions when solving for the wavefunctions and the electrostatic potential using \cref{eqn:GHEP,eqn:poissonElFE}, additionally we impose the constraint that the mean electrostatic potential in a periodic unit-cell is zero to obtain a unique solution for the electrostatic potential in \cref{eqn:poissonElFE}. Periodic boundary conditions are used in \qe{} for these calculations.
\subsubsection{Accuracy benchmarking}
In our accuracy benchmarking studies of periodic systems reported here comparing \ncsocfe{} with \qe{}, we use a kinetic energy cutoff (\texttt{ecutwfc}) of 75Ha in \qe{} while a polynomial order of 6 with a finite-element mesh size of 0.8 Bohr is used in our method.
\paragraph{Energetics:} Table~\eqref{tab:intEnergyper} shows the comparison of total internal energies with \qe{} for the relaxed, periodic face-centred-cubic GaAs primitive unit-cell that display spin-orbit interaction and cubic MnSi primitive unit-cell the exhibit noncollinear magnetism.  In both \qe{} and \ncsocfe{}, shifted $8 \times 8 \times 8$ and $3 \times 3 \times 3$ Monkhorst-Pack k-point grids are used to sample the Brillouin zone for GaAs and MnSi unit-cells, respectively. From the \cref{tab:intEnergyper}, we observe an excellent match between \ncsocfe{} and \qe{}. 
\begin{table}[!ht]
    \makegapedcells
    \setlength\tabcolsep{2pt}
    \begin{tabularx}{\linewidth}{|Y|Y|Y|Y|}
    \Xhline{2pt}
    System & QE (Ha) & \ncsocfe{} (Ha) & Error (Ha/atom)\\
    \Xhline{2pt}
    GaAs   &-182.528415 &-182.528306 &5.5E-05\\
    MnSi   &-452.671967 &-452.671811 &1.9E-05\\
    \Xhline{2pt}
    \end{tabularx}
    \caption{\label{tab:intEnergyper}Accuracy benchmarks for total internal energy.}
\end{table}
Additionally, we compare the volume integral of the magnitude of magnetization density for the case of MnSi primitive unit-cell in \cref{tab:absMagMnSi} and observe a close match between \ncsocfe{} and \qe{}.
\begin{table}[!ht]
    \makegapedcells
    \setlength\tabcolsep{2pt}
    \begin{tabularx}{\linewidth}{|Y|Y|Y|Y|}
    \Xhline{2pt}
    System & \qe{} (Bohr Magneton) & \ncsocfe{} (Bohr Magneton) & Error (Bohr Magneton)\\
    \Xhline{2pt}
    MnSi   & 4.6941&4.6902 &3.9E-03\\
    \Xhline{2pt}
    \end{tabularx}
    \caption{\label{tab:absMagMnSi}Accuracy benchmarks for volume integral of the magnitude of magnetization density.}
\end{table}
\paragraph{Magneto-Crystalline Anisotropy}
We also benchmark the magnetocrystalline anisotropy energy for the tetragonal bulk transition metal alloy FePt, which is known to exhibit magnetocrystalline anisotropy~\cite{AyazKhan2016MagnetocrystallineView,Blanco-Rey2019ValidityAnisotropy}. To this end, in both \qe{} and \ncsocfe{} we utilize a $16\times16\times12$ Monkhorst-Pack grid~\cite{Monkhorst1976SpecialIntegrations} and compute the energy difference for the cases with the magnetization density pointing along the z-axis ($E_z$) and with the magnetization density pointing along the x-axis ($E_x$)
\begin{table}[!ht]
    \makegapedcells
    \setlength\tabcolsep{2pt}
    \begin{tabularx}{\linewidth}{|Y|Y|Y|Y|}
    \Xhline{2pt}
    System & \qe{} (meV) & \ncsocfe{} (meV) & Error (meV)\\
    \Xhline{2pt}
    FePt   & 2.747 & 2.743 &4.1E-03\\
    \Xhline{2pt}
    \end{tabularx}
    \caption{\label{tab:mcaFePt}Accuracy benchmarks for the magneticrystalline anisotropy energy.}
\end{table}

\begin{figure}[!htb]
    \includegraphics[width=\linewidth]{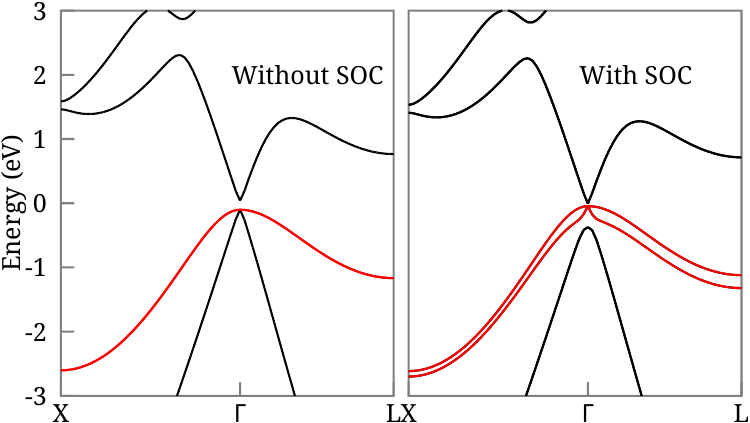}
    \caption{Bandstructure of GaAs with and without SOC, the bands shown in red serve as an example for band splitting}\label{fig:bands}
\end{figure}

\paragraph{Band-structure:} 
We also compute the band structure of GaAs, which is known to exhibit band-splitting due to SOC. To this end, we compute the ground-state electron density for the structures obtained from the Materials-Project database~\cite{Jain2013Commentary:Innovation} using the SCF procedure with a $8\times8\times8$ Monkhorst-Pack grid~\cite{Monkhorst1976SpecialIntegrations} for Brillouin zone sampling followed by a non-SCF calculation to obtain the eigenvalues along the chosen high-symmetry path. \Cref{fig:bands} shows the band structure of GaAs with and without SOC obtained from our method. 
We see that the band structure obtained from our \ncsocfe{} calculations demonstrates the splitting of the heavy hole band (shown in red). At the $\Gamma$-point, we see that SOC splits the triply degenerate valence states into a doubly degenerate state and a non-degenerate state. This is consistent with existing literature~\cite{Luo2009Full-ZoneGaSb,Christensen1984ElectronicStrain}.

\subsubsection{Performance studies}
\begin{table*}[t]
    \makegapedcells
    \setlength\tabcolsep{2pt}
     \begin{tabularx}{\linewidth}{|Y|Y|Y|Y|Y|Y|Y|Y|}
        \Xhline{2pt}
        Supercell size &Atoms & Electrons&\qe{} \# of basis fns.  & \qe{} Nurion CPU Node-hrs ($\eta_{\text{c}}$) & \ncsocfe{} \# of basis fns. & \ncsocfe{} Nurion CPU Node-hrs ($\eta_{\text{c}}$) & \ncsocfe{} Frontier GPU Node-hrs ($\eta_{\text{g}}$)\\
        \Xhline{2pt}
        6x6x1 & 288&2736 & 322917 & 0.8 & 4351250 & 2.5 & 0.04 \\
         \hline
         8x8x1 & 512&4864 & 574029 &5.7  & 7722450 & 8.4 & 0.2\\
         \hline
         10x10x1 & 800&7600 & 896981 & 26.5 & 12400200 & 26.0 & 0.6 \\
         \hline
          12x12x1 & 1152&10944 & 1291453 & 132.4 & 17760800 & 71.4 & 1.2 \\
         \hline
          14x14x1 & 1568&14896 & 1757965 & --\footnote{We were unable to perform this computation due to memory limitations on KNL architectures} & 24081800 & 154.4 & 2.8 \\
          \Xhline{2pt}
    \end{tabularx}
    \caption{Computational cost ($\eta$) comparison between \ncsocfe{} and \qe{} in node-hrs (discretization error $\sim 10^{-4}$~Ha/atom). $\eta$ is computed as the product of the minimum number of nodes required to fit a given material system and the average wall time taken per SCF iteration. \textbf{Case Study}: MnSi periodic supercells}
    \label{tab:computecostmnsi}
\end{table*}

We now discuss the performance of \ncsocfe{} with
\qe{} by comparing the average CPU time per SCF iteration in terms of computational node-hrs ($\eta_c$) with increasing system sizes of a periodic system solved to a similar
level of accuracy. Particularly, we consider MnSi supercells of increasing sizes. The Skyrmion radius in MnSi is $\sim 10$nm, and any computational study of Skyrmions in MnSi requires large supercells. To this end, periodic pseudopotential DFT calculations involving noncollinear magnetism and spin-orbit coupling
terms are conducted on MnSi supercells of sizes $6 \times 6 \times 1$ (2736 electrons), $8 \times 8 \times 1$ (4864 electrons), $10 \times 10 \times 1$ (7600 electrons), $12 \times 12 \times 1$ (10944 electrons), and $14 \times 14 \times 1$ (14896 electrons). \Cref{tab:computecostmnsi} summarizes the configurations and number of atoms considered in increasing supercell size.

We employ the PBE functional~\cite{PhysRevLett.77.3865} for the exchange-correlation and fully-relativistic optimized norm-conserving pseudopotentials (ONCV)~\cite{Hamann2013OptimizedPseudopotentials} in the Pseudo-dojo database~\cite{vanSetten2018TheTable}. In order to evaluate the performance in a consistent manner, we choose the basis set parameters such that the error in the ground-state energy due to the discretization is $\sim 10^{-4}$Ha/atom. To this end, we use a kinetic energy cutoff (\texttt{ecutwfc}) of 45Ha in \qe{} and a polynomial order of 7 with a mesh size of 1.2 Bohr in our method.  We perform a noncollinear calculation with SOC in both our method and \qe~using a 1x1x3 Monkhorst-Pack grid~\cite{Monkhorst1976SpecialIntegrations} for Brillouin zone sampling employing periodic boundary conditions in all three directions.

\Cref{tab:computecostmnsi} shows the average CPU time per SCF iteration in terms of computational node-hrs ($\eta_c$) and the number of basis functions in \ncsocfe{} and \qe{} for various sizes of MnSi supercells considered in this study. We find that for system sizes $\sim$8000 electrons and above, \ncsocfe{} becomes more efficient than \qe{} with increasing system size on CPUs. From the above table, the computational complexity in the regime of $\sim3000-15000$ electrons for \ncsocfe{} is estimated to be around $\order{(N_e^{2.3})}$ while in the case of \qe{}, it is around $\order{(N_e^{3.3})}$. The higher computational complexity of \qe{} is attributed to the requirement of a large number of processors to meet the peak memory requirement that affects the scalability of \qe{}. The efficient parallel scalability of our methods provides the necessary advantage in terms of computational node-hrs as the system size increases. Furthermore, \cref{tab:computecostmnsi} also reports the average time per SCF iteration in terms of GPU node-hrs ($\eta_g)$, and we find $\sim$ 40$\times-50\times$ Nurion CPU node-hr to Frontier GPU node-hr speedups, highlighting the significant advantage of our computational method on GPUs.
\subsubsection{Scalability}
\begin{figure}[!ht]
    \includegraphics[width=\linewidth]{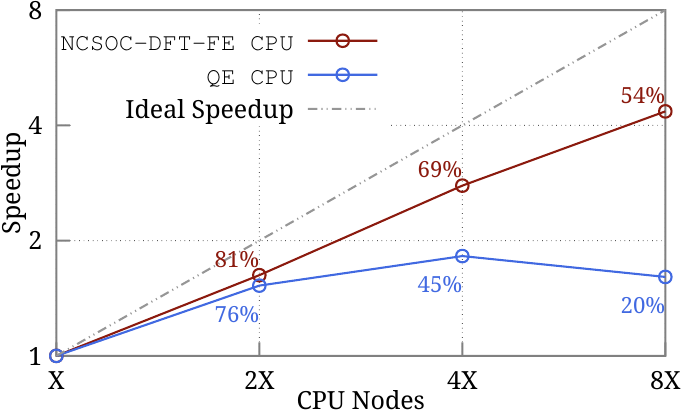}
    \caption{Relative speed ups of wall-time per SCF iteration per k-Point for the 8x8x1 supercell of MnSi with 4864 electrons for varying number of nodes on Nurion. \ncsocfe{}: The value of X is 100, and the total number of DoFs is 7722450. The number of band groups at X, 2X, 4X, and 8X is 1, 1, 2, and 4, respectively. \qe{}: The value of X is 16, and the total number of plane-waves is 574029. The number of band groups at X, 2X, 4X, and 8X is 1, 1, 2, and 2 respectively.}\label{fig:mnsiScalCPU}
\end{figure}

\begin{figure}[!hbt]
    \includegraphics[width=\linewidth]{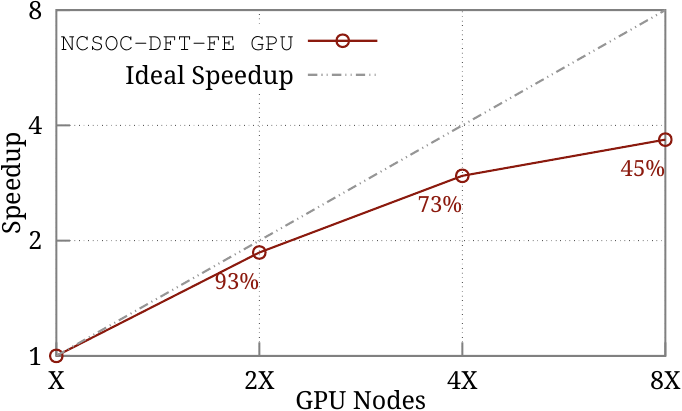}
    \caption{Relative speed ups of wall-time per SCF iteration per k-Point for the 8x8x1 supercell of MnSi with 4864 electrons for varying number of nodes on Frontier. \ncsocfe{}: The value of X is 8, and the total number of DoFs is 7722450.}\label{fig:mnsiScalGPU}
\end{figure}
\begin{table}[!b]
    \makegapedcells
    \setlength\tabcolsep{2pt}
    \begin{tabularx}{\linewidth}{|Y|Y|Y|}
        \Xhline{2pt}
        Method & \# of nodes & Minimum Walltime per SCF  ($\tau^\text{min}$)\\
        \Xhline{2pt}
       \qe{} Nurion CPUs& 64& 709\\
       \hline
       \ncsocfe{} Nurion CPUs & 800&69\\
       \hline
       \ncsocfe{} Frontier GPUs & 64&21\\
        \Xhline{2pt}
    \end{tabularx}
    \caption{Computational cost ($\tau^\text{min}$) comparison between \ncsocfe{} and \qe{} in terms of minimum walltime per scf per k-point (discretization error $\sim 10^{-4}$~Ha/atom). \textbf{Case Study}: 8x8x1 supercell of MnSi}
    \label{tab:computecostmnsiminwall}
\end{table}

We now demonstrate the parallel strong scaling of our implementation in the case of periodic $8 \times 8 \times 1$ supercell on both multi-node CPUs and GPUs. The MnSi supercell considered in this study involves around 7.7 million degrees of freedom (number of grid points/FE basis functions). \Cref{fig:mnsiScalCPU} compares the scaling behavior of \ncsocfe{} with \qe{} on Nurion KNL CPU nodes and \cref{fig:mnsiScalGPU} demonstrates the scalability of \ncsocfe{} on Frontier AMD GPUs, with the lowest number of nodes chosen in each case to be the minimum number of nodes that the problem could fit on. In both approaches, the discretization parameters are chosen such that the discretization error in ground-state energy is $\order{(10^{-4})}$ as in previous performance studies. As evident from the figure, we observe a scaling efficiency of $\sim$54 \%  even on 800 KNL compute nodes and an efficiency of $\sim$45 \% on 64 GPU nodes. Similar to the semi-periodic case of WTe$_2$, we find the relative speedups obtained in \qe{} are of a limited range, with speedups flattening off beyond a certain number of nodes. To this end, a comparison of minimum wall times per SCF iteration per k-point between the two approaches for this medium-scale system containing 4864 electrons shows a massive 11 $\times$ gains over \qe{} on multi-node CPU nodes and more than 30 $\times$ gains on multi-node GPU nodes, underscoring the excellent parallel scalability of our numerical implementation. We note that these gains will increase with an increase in system sizes.

\section{Summary}\label{sec:concl}
In this work, we present a computationally efficient and systematically convergent real-space finite-element based methodology for large-scale pseudopotential density functional theory (DFT) calculations incorporating noncollinear magnetism and spin-orbit coupling effects. The development of the proposed approach is based on the following key ideas. First, we deduced the finite-element (FE) discretized governing equations involving 2-component spinors wherin we have made use of a local real-space formulation to evaluate the electrostatic potential. We have also devised a numerical strategy extending the ideas of the White-Bird approach to evaluate the GGA exchange-correlation potential for noncollinear magnetism within the FE framework of DFT involving $C^{0}$ basis functions. This strategy mitigated the need to deal with the gradient of the magnetization direction, whose integral is unbounded as the magnetization tends to zero. Subsequently, we developed efficient methods that exploit the sparsity of local and non-local parts of the FE discretized Hamiltonian matrix to compute the action of this matrix on a trial subspace of vectors. These methods were leveraged alongside a self-consistent field (SCF) iteration approach with the Chebyshev filtered subspace iteration procedure to solve the underlying FE discretized generalized eigenvalue problem based on a two-grid strategy. Furthermore, we introduced a unified approach to compute atomic forces and unit-cell stresses in the presence of noncollinear magnetism and spin-orbit coupling. This approach, based on the configurational force method, evaluates the directional derivative of a generalized energy functional with the a stationary point at the minima of the Kohn-Sham functional.

We validated the accuracy of the proposed method against a plane-wave implementation of noncollinear magnetism and spin-orbit interaction across various representative benchmark systems, including non-periodic, semi-periodic, and fully periodic systems. Our proposed methodology(\ncsocfe{}) showed excellent agreement with the plane-wave calculations for several key metrics, including ground-state energies, vertical ionization potentials, magnetic anisotropy energies, band structures, and spin-textures. Furthermore, compared to the state-of-the-art plane-wave implementation, the efficient computational strategies employed in \ncsocfe{} enabled a $2\times$ reduction in computational cost in material systems involving ~15000-20000 electrons on CPUs. Additionally, our method achieves substantial speed ups ($\sim$ 8$\times-11\times$) in terms of minimum wall-time compared to state-of-the-art plane-wave code (\qe) for both semi-periodic and non-periodic systems. Furthermore, the proposed computational strategies exhibit excellent parallel scaling efficiency (up to $\sim63\%$) even on 800 KNL compute nodes. We have also demonstrated the performance of our methodology on the Frontier exascale system wherein we have achieved $\sim 50x$ reduction in node-hrs compared to the KNL CPU architecture. We also achieve parallel scaling efficiency of up to $\sim 45\%$ on 64 nodes on Frontier.

The proposed work, \ncsocfe{}, takes advantage of the compactly supported nature of finite-element (FE) basis functions and their adaptive resolution, enabling efficient use of modern supercomputing architectures, resulting in a substantial computational advantage over the current state-of-the-art plane-wave-based implementations for medium to large-scale material systems. Thus, the proposed methodology allows for studying noncollinear magnetism and spin-orbit coupling effects on larger length scales from first principles, significantly enhancing the scope of \emph{ab-initio} computations for systems where such effects play a crucial role. 

We note that the current implementation does not utilize the recently proposed matrix-free algorithms for matrix multivector products~\cite{Panigrahi2024FastSystems}, which take advantage of the tensor-structured nature of the FE basis and show great promise in further accelerating real-space FE-based density functional theory computations. Hence, as part of future work, we aim to extend the matrix-free algorithms to the two-component complex spinor framework required to incorporate noncollinear magnetism and spin-orbit coupling effects in DFT, which would further reduce the computational cost of our method. We also aim to extend our proposed methodology and incorporate noncollinear magnetism and spin-orbit coupling effects within the framework of the recently introduced finite-element based projector-augmented wave method~\cite{Ramakrishnan2024FastMethod} which has been shown to require considerably fewer number of basis functions when compared to the ONCV framework. Finally, we aim to utilize our proposed methodologies to solve large-scale problems of scientific interest involving SOC and noncollinear magnetism employing periodic/semi-periodic/fully non-periodic boundary conditions from an \emph{ab-initio} perspective.


\section{Acknowledgements}
 The authors gratefully acknowledge the seed grant from the Indian Institute of Science and the SERB Startup Research Grant from the Department of Science and Technology India (Grant Number: SRG/2020/002194). The research used the resources of PARAM Pravega at the Indian Institute of Science, supported by the National Supercomputing Mission (NSM). The research also used resources from KIST Supercomputer Nurion in South Korea for all the CPU calculations. This research also used GPU resources from the Oak Ridge Leadership Computing Facility at the Oak Ridge National Laboratory, supported by the Office of Science of the U.S. Department of Energy under Contract No. DE-AC05-00OR22725. N.K. would like to acknowledge the Prime Minister Research Fellowship(PMRF) from the Ministry of Education India and the Council of Scientific and Industrial Research (CSIR) fellowship for financial support. P.M. acknowledges Google India Research Award 2023 and Indo-Korean Science and Technology Center Bengaluru for financial support during the course of this work

\begin{widetext}
\revappendix
\section{Stationary properties of the generalized functional}\label[appendix]{sec:appendixGenFunc}
Consider the generalized functional
\begin{multline}
    \widetilde{E}[\widetilde{V},\widetilde{\bB},\widetilde{\rho},\widetilde{\bmm},\widetilde{f}_{n\bk}]=\sum_{n}\fint_{BZ}\widetilde{f}_{n\bk}\epsilon_{n\bk}[\widetilde{V},\widetilde{\bB}]\dk-\int \left(\widetilde{V}\widetilde{\rho}+\widetilde{\bB}\cdot\widetilde{\bmm}\right)\dr
    +E_\text{el}[\widetilde{\rho},\bR]+E_\text{xc}[\widetilde{\rho},\widetilde{\bmm}]\\+E_\text{ent}[\widetilde{f}_{n\bk}]+\mu\left(N_e-\sum_{n}\fint_{BZ}\widetilde{f}_{n\bk}\dk\right)
\end{multline}
The functional derivatives of this functional can be written as
\begin{align}
    \funcder{\widetilde{E}}{\widetilde{V}}&=\rho[\widetilde V,\widetilde \bB]-\widetilde\rho\label{eqn:genFuncDerRho}\\
    \funcder{\widetilde{E}}{\widetilde\bB}&=\bmm[\widetilde V,\widetilde \bB]-\widetilde\bmm\label{eqn:genFuncDerMag}\\
    \funcder{\widetilde{E}}{\widetilde\rho}&=V[\widetilde \rho,\widetilde \bmm]-\widetilde V\label{eqn:genFuncDerV}\\
    \funcder{\widetilde{E}}{\widetilde\bmm}&=\bB[\widetilde \rho,\widetilde \bmm]-\widetilde\bB\label{eqn:genFuncDerB}\\
    \funcder{\widetilde{E}}{\widetilde f_{n\bk}}&=\epsilon_{n\bk}[\widetilde V,\widetilde \bB]-\mu+k_BT\ln{\frac{\widetilde f_{n\bk}}{1-\widetilde f_{n\bk}}}\label{eqn:genFuncDerOcc}
\end{align}
where we have defined $\rho\left[\widetilde V,\widetilde\bB\right]$ and $\bmm\left[\widetilde V,\widetilde\bB\right]$ as the total charge and magnetization densities computed using the wavefunctions obtained as solution of the eigenvalue problem defined by \cref{eqn:genKSProblem} using \cref{eqn:rho,eqn:mag}. We also define $V\left[\widetilde\rho,\widetilde\bmm\right]$ and $\bB\left[\widetilde\rho,\widetilde\bmm\right]$ as 
\begin{align}
    V\left[\widetilde\rho,\widetilde\bmm\right]&=V_\text{el}\left[\widetilde\rho\right]+V_\text{xc}\left[\widetilde\rho,\widetilde\bmm\right] \\ \bB\left[\widetilde\rho,\widetilde\bmm\right]&=\bB_\text{xc}\left[\widetilde\rho,\widetilde\bmm\right]
\end{align}
where $V_\text{xc}\left[\widetilde\rho,\widetilde\bmm\right]$ and $\bB_\text{xc}\left[\widetilde\rho,\widetilde\bmm\right]$ are given by
\begin{align}
    V_\text{xc}\left[\widetilde\rho,\widetilde\bmm\right]&=\frac{\delta E_\text{xc}}{\delta\rho}\biggr|_{\substack{\rho=\widetilde\rho\\\bmm=\widetilde\bmm}}&\bB_\text{xc}\left[\widetilde\rho,\widetilde\bmm\right]&=\frac{\delta E_\text{xc}}{\delta\bmm}\biggr|_{\substack{\rho=\widetilde\rho\\\bmm=\widetilde\bmm}}
\end{align}
and $V_\text{el}\left[\widetilde\rho\right]$ is given as the solution of
\begin{align}
    -\nabla^2V_\text{el}=4\pi\left(\widetilde\rho+b\right)
\end{align}

Setting the functional derivatives in \cref{eqn:genFuncDerRho,eqn:genFuncDerMag,eqn:genFuncDerV,eqn:genFuncDerB,eqn:genFuncDerOcc} to zero and after straightforward algebraic manipulations we obtain
\begin{align}
    \rho[V[\widetilde\rho,\widetilde\bmm],\bB[\widetilde\rho,\widetilde\bmm]]&=\widetilde\rho\\
    \bmm[V[\widetilde\rho,\widetilde\bmm],\bB[\widetilde\rho,\widetilde\bmm]]&=\widetilde\bmm
\end{align}
Or equivalently
\begin{align}
    F\left[\left(\widetilde\rho,\widetilde\bmm\right)\right]=\left(\widetilde\rho,\widetilde\bmm\right)
\end{align}
which is the solution to the Kohn-Sham eigenproblem defined by \cref{eqn:nlEig,eqn:KSMap}. Thus we can conclude that the generalized functional defined in \cref{eqn:genEnergyFunc} has a stationary point at the solution to the Kohn-Sham eigenvalue problem.
\section{Gateaux derivatives of the generalized functional}\label[appendix]{sec:appendixGatDer}
In order to compute the configurational force on a system due to a perturbation $\btau^\varepsilon$ we need to evaluate
\begin{align}
    \frac{d\widetilde{E}_S^{\varepsilon}}{d\varepsilon}\biggl|_{\varepsilon=0}
\end{align}

To this end we first write the parametrized functional $\widetilde{E}_S^\varepsilon$ in the perturbed space as
\begin{multline}
    \widetilde{E}_S^\varepsilon[V^\varepsilon,\bB^\varepsilon,\rho^\varepsilon,\bmm^\varepsilon,{f}_{n\bk^\varepsilon}^\varepsilon]=\sum_{n}\fint_{BZ^\varepsilon}f_{n\bk^\varepsilon}^\varepsilon\epsilon_{n\bk^\varepsilon}^\varepsilon[V^\varepsilon,\bB^\varepsilon]\dk^\varepsilon-\int \left(V^\varepsilon\rho^\varepsilon+\bB^\varepsilon\cdot\bmm^\varepsilon\right)\dr^\varepsilon
    +E_\text{el}^\varepsilon[\rho^\varepsilon,\bR^\varepsilon]+E_\text{xc}^\varepsilon[\rho^\varepsilon,\bmm^\varepsilon]\\+E_\text{ent}^\varepsilon[f_{n\bk^\varepsilon}^\varepsilon]+\mu\left(N_e-\sum_{n}\fint_{BZ^\varepsilon}f_{n\bk^\varepsilon}^\varepsilon\dk^\varepsilon\right)
\end{multline}
where $V^\varepsilon,\bB^\varepsilon,\rho^\varepsilon,\bmm^\varepsilon$ and ${f}_{n\bk}^\varepsilon$ represent the solutions of the statationary point in the perturbed space and consequently they satisfy the Euler-lagrange equations obtained by setting \cref{eqn:genFuncDerRho,eqn:genFuncDerMag,eqn:genFuncDerV,eqn:genFuncDerB,eqn:genFuncDerOcc} to zero. Note that $\btau^{\varepsilon=0}=\mathbb{I}$ and consequently we drop the superscript of $\varepsilon$ when $\varepsilon=0$ for notational convenience. The derivative can now be written as
\begin{multline}
    \derveps{\widetilde{E}_S^\varepsilon[V^\varepsilon,\bB^\varepsilon,\rho^\varepsilon,\bmm^\varepsilon,{f}_{n\bk}^\varepsilon]}\biggl|_{\varepsilon=0}=\parder{\widetilde{E}_S^\varepsilon[V,\bB,\rho,\bmm,{f}_{n\bk}]}{\varepsilon}\biggl|_{\varepsilon=0}+\int\dr^\varepsilon\funcder{\widetilde{E}_S^\varepsilon}{\rho^\varepsilon(\br^\varepsilon)}\dervepsZ{\rho^\varepsilon(\br^\varepsilon)}+\int\dr^\varepsilon\funcder{\widetilde{E}_S^\varepsilon}{\bmm^\varepsilon(\br^\varepsilon)}\cdot\dervepsZ{\bmm^\varepsilon(\br^\varepsilon)}\\+\int\dr^\varepsilon\funcder{\widetilde{E}_S^\varepsilon}{V^\varepsilon(\br^\varepsilon)}\dervepsZ{V^\varepsilon(\br^\varepsilon)}+\int\dr^\varepsilon\funcder{\widetilde{E}_S^\varepsilon}{\bB^\varepsilon(\br^\varepsilon)}\cdot\dervepsZ{\bB^\varepsilon(\br^\varepsilon)}+\fint_{BZ}\dk^\varepsilon\funcder{\widetilde{E}_S^\varepsilon}{f_{n\bk^\varepsilon}^\varepsilon}\dervepsZ{f_{n\bk^\varepsilon}^\varepsilon}
\end{multline}
we note that at the stationary point, all the functional derivatives (\cref{eqn:genFuncDerRho,eqn:genFuncDerMag,eqn:genFuncDerV,eqn:genFuncDerB,eqn:genFuncDerOcc}) are zero, consequently we can write
\begin{equation}
    \derveps{\widetilde{E}_S^\varepsilon[V^\varepsilon,\bB^\varepsilon,\rho^\varepsilon,\bmm^\varepsilon,{f}_{n\bk}^\varepsilon]}\biggl|_{\varepsilon=0}=\parder{\widetilde{E}_S^\varepsilon[V,\bB,\rho,\bmm,{f}_{n\bk}]}{\varepsilon}\biggl|_{\varepsilon=0}
\end{equation}
where $\widetilde{E}_S^\varepsilon[V,\bB,\rho,\bmm,{f}_{n\bk}]$ can be written as
\begin{multline}
    \widetilde{E}_S^\varepsilon[V,\bB,\rho,\bmm,{f}_{n\bk}]=\sum_{n}\fint_{BZ^\varepsilon}f_{n\bk}\epsilon_{n\bk^\varepsilon}^\varepsilon[V,\bB]\dk^\varepsilon-\int \left(V\rho+\bB\cdot\bmm\right)\dr^\varepsilon
    +E_\text{el}^\varepsilon[\rho,\bR^\varepsilon]+E_\text{xc}^\varepsilon[\rho,\bmm]\\+E_\text{ent}^\varepsilon[f_{n\bk}]+\mu\left(N_e-\sum_{n}\fint_{BZ^\varepsilon}f_{n\bk}\dk^\varepsilon\right)
\end{multline}
We can now transform the integrals in the above equation into integrals over the unperturbed space, allowing us to write
\begin{multline}
    \widetilde{E}_S^{\varepsilon}=\sum_{n}\fint_{BZ}f_{n\bk}\epsilon_{n\bk^\varepsilon}^\varepsilon[V,\bB]\dk-\int \left(V\rho+\bB\cdot\bmm\right)\detb{\parder{\br^\varepsilon}{\br}}\dr
    +E_\text{el}^\varepsilon[\rho,\bR^\varepsilon]+E_\text{xc}^\varepsilon[\rho,\bmm]\\+E_\text{ent}^\varepsilon[f_{n\bk}]+\mu\left(N_e-\sum_{n}\fint_{BZ}f_{n\bk}\dk\right)
\end{multline}
The required partial derivative can now be written as
\begin{multline}
    \pardervepsZ{\widetilde{E}_S^{\varepsilon}}=\sum_{n}\fint_{BZ}\dk f_{n\bk}\pardervepsZ{\epsilon_{n\bk^\varepsilon}^\varepsilon[V,\bB]}-\int\dr \left(V\rho+\bB\cdot\bmm\right)\pardervepsbZ{\detb{\parder{\br^\varepsilon}{\br}}}+\pardervepsZ{E_\text{el}^\varepsilon[\rho,\bR^\varepsilon]}\\+\pardervepsZ{E_\text{xc}^\varepsilon[\rho,\bmm]}\label{eqn:parderGenFunc}
\end{multline}
For the evaluation of the partial derivative of the term corresponding to the total electrostatic energy we refer to \cite{Das2022DFT-FEDiscretization,Motamarri2018ConfigurationalTheory}. We now discuss the evaluation of the rest of the terms.
\subsection{Gateaux derivative of the eigenvalue}
Consider the first term in \cref{eqn:parderGenFunc}, we need to evaluate the derivative of the eigenvalues in the perturbed space
\begin{align}
    \parderveps{\epsilon_{n\bk^\varepsilon}^\varepsilon[V,\bB]}=\parderveps{\bra{\bu_{n\bk^\varepsilon}^\varepsilon}\mathcal{H}^{\bk^\varepsilon}_\varepsilon[V,\bB]\ket{\bu_{n\bk^\varepsilon}^\varepsilon}_\varepsilon}
\end{align}
We can now invoke the Hellmann–Feynman theorem 
\begin{align}
    \parderveps{\epsilon_{n\bk^\varepsilon}^\varepsilon[V,\bB]}&=\bra{\bu_{n\bk^\varepsilon}^\varepsilon}\parderveps{\mathcal{H}^{\bk^\varepsilon}_\varepsilon[V,\bB]}\ket{\bu_{n\bk^\varepsilon}^\varepsilon}_\varepsilon
\end{align}
We can now write
\begin{align}
    \parderveps{\mathcal{H}^{\bk^\varepsilon}_\varepsilon[V,\bB]}&=\parderveps{\mathcal{H}^{\bk^\varepsilon,loc}_\varepsilon[V,\bB]}+\parderveps{\mathcal{H}^{\bk^\varepsilon,nloc}_\varepsilon}
\end{align}

Consequently for the local part of the Hamiltonian we can write
\begin{multline}
    \bra{\bu_{n\bk}}\pardervepsZ{\mathcal{H}^{\bk^\varepsilon,loc}_\varepsilon[V,\bB]}\ket{\bu_{n\bk}}=\parderveps{}\Biggl\{\int\Biggl(-\frac{\bu_{n\bk}^{ \dagger} \nabla_\varepsilon^2\bu_{n\bk}}{2}-\bu_{n\bk}^{\dagger}\iu \bk^\varepsilon\cdot\nabla_\varepsilon\bu_{n\bk}+\frac{\abs{\bk^\varepsilon}^2}{2}\bu_{n\bk}^{\dagger}\bu_{n\bk}
    +(V+V_\text{loc}^\varepsilon-V_\text{self}^\varepsilon)\bu_{n\bk}^{\dagger}\bu_{n\bk}\\+\bu_{n\bk}^{\dagger}\bB \cdot\bsigma\bu_{n\bk}\Biggr)\dr\Biggr\}\Bigg|_{\varepsilon=0}
\end{multline}
Upon further simplification, this results in
    \begin{multline}
        \bra{\bu_{n\bk}}\parderveps{\mathcal{H}^{\bk^\varepsilon,loc}_\varepsilon[V,\bB]}\ket{\bu_{n\bk}}=\int\Biggl( \frac{1}{2}\left(\nabla\bu_{n\bk}^\dagger\otimes\nabla\bu_{n\bk}+\nabla\bu_{n\bk}\otimes\nabla\bu_{n\bk}^\dagger-\iu\bu_{n\bk}^\dagger\nabla\bu_{n\bk}\otimes\bk\right):\pardervepsb{\parder{\br}{\br^\varepsilon}}\\+\frac{1}{2}\left(\nabla\bu_{n\bk}^\dagger\cdot\nabla\bu_{n\bk}+\bu_{n\bk}^\dagger\nabla^2\bu_{n\bk}\right)\pardervepsb{\det{\parder{\br^\varepsilon}{\br}}}\\
        -\iu\bu_{n\bk}^\dagger\parderveps{\bk^\varepsilon}\cdot\nabla\bu_{n\bk}+\frac{1}{2}\pardervepsb{\abs{\bk^\varepsilon}^2}\bu_{n\bk}^\dagger\bu_{n\bk}+\parderveps{V_\text{loc}^\varepsilon}\bu_{n\bk}^\dagger\bu_{n\bk}-\parderveps{V_\text{self}^\varepsilon}\bu_{n\bk}^\dagger\bu_{n\bk}\Biggr)\dr\label{eqn:GateauxHLoc}
    \end{multline}
In a similiar manner, for the non-local part we can write
\begin{multline}
    \bra{\bu_{n\bk}}\pardervepsZ{\mathcal{H}^{\bk^\varepsilon,nloc}_\varepsilon[V,\bB]}\ket{\bu_{n\bk}}=\sum_{a\in \Omega_p}\sum_{\rchi \rchi^\prime}\sum_q \parderveps{}\int_{\Omega_p}\biggl(\bu_{n\bk}^\dagger(\br) e^{-\iu\bk^\varepsilon\cdot(\br^\varepsilon-\bL_q^\varepsilon)}p_\rchi^a(\br^\varepsilon-\bL_q^\varepsilon-\bR_a^\varepsilon)\biggr)\dr\\\bD^{\gamma_a,\rchi,\rchi^\prime}\int_{\Omega_p} \biggl(\sum_{q^\prime}e^{\iu\bk^\varepsilon\cdot({\br^\prime}^\varepsilon-\bL_{q^\prime}^\varepsilon)}p^a_{\rchi^\prime}({\br^\prime}^\varepsilon-\bL_{q^\prime}^\varepsilon-\bR_a^\varepsilon)\bu_{n\bk}(\br^\prime)\biggr)\det{\parder{{\br^\prime}^\varepsilon}{\br^\prime}}\dr^\prime\label{eqn:GateauxHnLoc}
\end{multline}
We now note the following relations
\begin{align}
    \pardervepsbZ{\det{\parder{\br^\varepsilon}{\br}}}&=\bI:\nabla\bUpsilon\\
    \pardervepsbZ{\parder{\br}{\br^\varepsilon}}&=-\nabla\bUpsilon
\end{align}
We also note that the Laplacian of the wavefunctions can be written as
\begin{align}
    \frac{1}{2}\nabla^2u_{n\bk}=\Biggl(\frac{\abs{\bk}^2}{2}-\iu\bk\cdot\nabla+V_\text{loc}-V_\text{self}+V+\bB\cdot\bsigma+\mathcal{H}^{\bk,nloc}-\epsilon_{n\bk}\Biggr)u_{n\bk}
\end{align}
Combining \cref{eqn:GateauxHLoc,eqn:GateauxHnLoc} and using the above relations, we have
    \begin{align}
        \sum_{n}\fint_{BZ}\dk f_{n\bk}\pardervepsZ{\epsilon_{n\bk^\varepsilon}^\varepsilon[V,\bB]}=\int_{\Omega_p}\bE_1:\nabla\bUpsilon\dr+\text{F}^\text{psp,nloc}+\text{F}^{K}+\text{F}^\text{ext,corr}
    \end{align}
    where $\bE_1$ is given by
    \begin{multline}
        \bE_1=\Biggl(\sum_{n}\fint_{BZ} \frac{f_{n\bk}}{2}\left(\nabla\bu_{n\bk}^\dagger\cdot\nabla\bu_{n\bk}+\left(\abs{\bk}^2-\epsilon_{n\bk}\right)\bu_{n\bk}^\dagger \bu_{n\bk}-\iu \bu_{n\bk}^\dagger\bk\cdot\nabla \bu_{n\bk}\right)\dk+\left(V_\text{loc}-V_\text{self}\right)\rho+V\rho+\bB\cdot\bmm\Biggr)\bI\\-\sum_{n}\fint_{BZ} \frac{f_{n\bk}}{2}\left(\nabla\bu_{n\bk}^\dagger\otimes\nabla\bu_{n\bk}+\nabla\bu_{n\bk}\otimes\nabla\bu_{n\bk}^\dagger-\iu\bu_{n\bk}^\dagger\nabla\bu_{n\bk}\otimes\bk\right)\dk
    \end{multline}
    The term $\text{F}^\text{psp,nloc}$ is given by $\text{F}^\text{psp,nloc}=\text{F}^\dagger_\text{nloc}+\text{F}_\text{nloc}$ where $\text{F}_\text{nloc}$ is given by
    \begin{multline}
        \text{F}_\text{nloc}=\sum_{a\in \Omega_p}\sum_{\rchi \rchi^\prime}\sum_q  \sum_{n}\fint_{BZ}\Biggl[\int_{\Omega_p}\biggl(\bu_{n\bk}^\dagger(\br) e^{-\iu\bk\cdot(\br-\bL_q)}p_\rchi^a(\br-\bL_q-\bR_a)\biggr)\dr\bD^{\gamma_a,\rchi,\rchi^\prime}\int_{\Omega_p} \biggl(\sum_{q^\prime}e^{\iu\bk\cdot({\br^\prime}-\bL_{q^\prime})}\\p^a_{\rchi^\prime}({\br^\prime}-\bL_{q^\prime}-\bR_a)\left(-\left(\bUpsilon(\br^\prime)-\bUpsilon\left(\bR_a+\bL_{q^\prime}\right)\right)\cdot\nabla\bu_{n\bk}(\br^\prime)-\iu\bk\cdot\bUpsilon(\bR_a)\bu_{n\bk}(\br^\prime)\right)\biggr)\dr^\prime\Biggr]\dk
    \end{multline}
    The term $\text{F}^K$ is given by
    \begin{multline}
        \text{F}^K=\sum_{n}\fint_{BZ}\int_{\Omega_p}\Biggl[-\iu\bu_{n\bk}^\dagger\pardervepsZ{\bk^\varepsilon}\cdot\nabla\bu_{n\bk}+\frac{1}{2}\pardervepsbZ{\abs{\bk^\varepsilon}^2}\bu_{n\bk}^\dagger\bu_{n\bk}\Biggr]\dr\dk\\+\sum_{a\in \Omega_p}\sum_{\rchi \rchi^\prime}\sum_q \sum_{n}\parderveps{}\Biggl\{\fint_{BZ}\Biggl[\int_{\Omega_p}\biggl(\bu_{n\bk}^\dagger(\br) e^{-\iu\bk^\varepsilon\cdot(\br-\bL_q)}p_\rchi^a(\br-\bL_q-\bR_a)\biggr)\dr\\\bD^{\gamma_a,\rchi,\rchi^\prime}\int_{\Omega_p} \biggl(\sum_{q^\prime}e^{\iu\bk^\varepsilon\cdot({\br^\prime}-\bL_{q^\prime})}p^a_{\rchi^\prime}({\br^\prime}-\bL_{q^\prime}-\bR_a)\bu_{n\bk}(\br^\prime)\biggr)\dr^\prime\Biggr]\dk\Biggr\}\Bigg|_{\varepsilon=0}
    \end{multline}
    Finally, the $\text{F}^\text{ext,corr}$ term is given by
    \begin{multline}
        \text{F}^\text{ext,corr}=\sum_{a\in\Omega_p}\sum_q\int_{\Omega_p}\rho\Biggl(\nabla V_\text{loc}^a(\br-\bR_a-\bL_q)\cdot\left(\bUpsilon(\br)-\bUpsilon(\bR_a+\bL_q)\right)-\nabla V_\text{self}^a(\br,\bR_a+\bL_q)\cdot\bUpsilon(\br)\\-\pardervepsZ{V_\text{self}^a(\br,b^a({\br^\prime}^\varepsilon-\bR_a^\varepsilon-\bL_q^\varepsilon))}\Biggr)\dr
    \end{multline}
\subsection{Gateaux derivative of the electrostatic energy}
In order to compute the Gateaux derivative of the total electrostatic energy, we follow the methodology prescribed in prior works~\cite{Motamarri2018ConfigurationalTheory,Das2022DFT-FEDiscretization,Rufus2022IonicBasis} and define the total electrostatic energy functional ($\widetilde{E}_\text{el}[\widetilde{V}_\text{el},\{\widetilde{V}_\text{self}^a\}_{a=1}^{N_a};\rho,\bR]$) as
\begin{multline}
    E_\text{el}[\rho,\bR]=\min_{\widetilde{V}_\text{self}^a\in H^1(\mathbb{R}^3)}\max_{\widetilde{V}_\text{el}\in H_\text{per}^1(\Omega_p)}\widetilde{E}_\text{el}[\widetilde{V}_\text{el},\{\widetilde{V}_\text{self}^a\}_{a=1}^{N_a};\rho,\bR]=\max_{\widetilde{V}_\text{el}\in H_\text{per}^1(\Omega_p)}\Biggl\{\int_{\Omega_p}\left[-\frac{1}{8\pi}\abs{\nabla \widetilde{V}_\text{el}}^2+(\rho+b(\br,\bR))\widetilde{V}_\text{el}\right]\dr\Biggr\}\\-\sum_{a\in \Omega_p}\max_{\widetilde{V}_\text{self}^a\in H^1(\mathbb{R}^3)}\Biggl\{\int_{\mathbb{R}^3}\left[-\frac{1}{8\pi}\abs{\nabla \widetilde{V}_\text{self}^a}^2+b^a(\br-\bR_a)\widetilde{V}_\text{self}^a\right]\dr\Biggr\}
\end{multline}
Consequently in the perturbed space we can write
\begin{multline}
    \widetilde{E}_\text{el}^\varepsilon[\widetilde{V}_\text{el}^\varepsilon,\{{\widetilde{V}_\text{self}^{a^\varepsilon}}\}_{a=1}^{N_a};\rho,\bR]=\int_{\Omega_p^\varepsilon}\left[-\frac{1}{8\pi}\abs{\nabla_\varepsilon \widetilde{V}_\text{el}^\varepsilon}^2+(\rho+b(\br^\varepsilon,\bR^\varepsilon))\widetilde{V}_\text{el}^\varepsilon\right]\dr^\varepsilon-\int_{\mathbb{R}^3}\left[-\frac{1}{8\pi}\abs{\nabla_\varepsilon \widetilde{V}_\text{self}^{a^\varepsilon}}^2+b^a(\br^\varepsilon-\bR_a^\varepsilon){\widetilde{V}_\text{self}^{a^\varepsilon}}\right]\dr^\varepsilon
\end{multline}
We now define $V_\text{el}^\varepsilon$ and $V_\text{self}^{a^\varepsilon}$ as the solutions of the above saddle-point problem in the perturbed space. This allows us to write the Gateaux dervative of the total electrostatic energy as
\begin{align}
    \pardervepsZ{E_\text{el}^\varepsilon[\rho,\bR^\varepsilon]}=\pardervepsZ{{E}_\text{el}^\varepsilon[{V}_\text{el},\{{{V}_\text{self}^{a}}\}_{a=1}^{N_a};\rho,\bR]}+\int\dr^\varepsilon\funcder{\widetilde{E}_\text{el}^\varepsilon}{V_\text{el}^\varepsilon}\dervepsZ{V_\text{el}^\varepsilon}+\int\dr^\varepsilon\funcder{\widetilde{E}_\text{el}^\varepsilon}{V_\text{self}^{a^\varepsilon}}\dervepsZ{V_\text{self}^{a^\varepsilon}}
\end{align}
We note that at the solution of the saddle-point problem the functional derivatives vanish, consequently we have
\begin{align}
    \pardervepsZ{E_\text{el}^\varepsilon[\rho,\bR^\varepsilon]}=\pardervepsZ{{E}_\text{el}^\varepsilon[{V}_\text{el},\{{{V}_\text{self}^{a}}\}_{a=1}^{N_a};\rho,\bR]}
\end{align}
Thus we have
\begin{multline}
    \pardervepsZ{E_\text{el}^\varepsilon[\rho,\bR^\varepsilon]}=\parderveps{}\Biggl\{\int_{\Omega_p}\biggl[-\frac{1}{8\pi}\abs{\nabla_\varepsilon \widetilde{V}_\text{el}}^2+(\rho+b(\br^\varepsilon,\bR^\varepsilon))\widetilde{V}_\text{el}\biggr]\det{\parder{\br^\varepsilon}{\br}}\dr\\-\int_{\mathbb{R}^3}\biggl[-\frac{1}{8\pi}\abs{\nabla_\varepsilon \widetilde{V}_\text{self}^{a}}^2+b^a(\br^\varepsilon-\bR_a^\varepsilon){\widetilde{V}_\text{self}^{a}}\biggr]\det{\parder{\br^\varepsilon}{\br}}\dr\Biggr\}
\end{multline}
Finally we can write
\begin{align}
    \pardervepsZ{E_\text{el}^\varepsilon[\rho,\bR^\varepsilon]}=\int_{\Omega_p}\bE_2:\nabla\bUpsilon\dr+\sum_{a\in \Omega_p}\int_{\mathbb{R}^3}\bE^a:\nabla\bUpsilon\dr+\text{F}^\text{sm}
\end{align}
where $\bE_2$ and $\bE^a$ are given by
\begin{align}
    \bE_2&=\left(-\frac{1}{8\pi}\abs{\nabla V_\text{el}}^2+\rho V_\text{el}\right)\bI+\frac{1}{4\pi}\nabla V_\text{el}\otimes \nabla V_\text{el}\\
    \bE^a&=\frac{1}{8\pi}\abs{\nabla V_\text{self}^a}^2\bI-\frac{1}{4\pi}\nabla V_\text{self}^a\otimes \nabla V_\text{self}^a
\end{align}
The term $\text{F}^\text{sm}$ is given by
\begin{align}
    \text{F}^\text{sm}&=\sum_{a\in \Omega_p}\sum_q\int_{\Omega_p}b^a(\br-\bR_a-\bL_q)\nabla V_\text{el}\cdot\left(\bUpsilon(\br)-\bUpsilon(\bR_a+\bL_q)\right)\dr-\sum_{a\in \Omega_p}\int_{\mathbb{R}^3}b^a(\br-\bR_a)\nabla V_\text{self}^a\cdot(\bUpsilon(\br)-\bUpsilon(\bR_a))\dr
\end{align}
\subsection{Gateaux derivative of the exchange-correlation energy}
We now compute the Gateaux derivative of the exchange-correlation term
\begin{align}
    \pardervepsZ{E_\text{xc}^\varepsilon[\rho,\bmm]}=\pardervepsbZ{\int_{\Omega_p}f_\text{xc}(\rho,\bmm,\nabla_\varepsilon\rho,\nabla_\varepsilon\bmm)\det{\parder{\br^\varepsilon}{\br}}\dr}
\end{align}
this results in 
\begin{align}
    \pardervepsZ{E_\text{xc}^\varepsilon[\rho,\bmm]}=\int_{\Omega_p}\bE_3:\nabla\bUpsilon\dr
\end{align}
where $\bE_3$ is given by
\begin{align}
    \bE_3=f_\text{xc}(\rho,\bmm,\nabla\rho,\nabla\bmm)\bI-\parder{f_\text{xc}}{\nabla\rho}\otimes\nabla\rho-\parder{f_\text{xc}}{\nabla \abs{\bmm}}\otimes\nabla \abs{\bmm}
\end{align}

The total configurational force is now given by
\begin{align}
    \frac{d\widetilde{E}_S^{\varepsilon}}{d\varepsilon}\biggl|_{\varepsilon=0}=\int_{\Omega_p}\bE:\nabla\bUpsilon\dr+\sum_{a\in \Omega_p}\int_{\mathbb{R}^3}\bE^a:\nabla\bUpsilon\dr+\text{F}^\text{psp,nloc}+\text{F}^{K}+\text{F}^\text{ext,corr}+\text{F}^\text{sm}
\end{align}
Where $\bE$ is given by $\bE=\bE_1+\bE_2+\bE_3-\rho V-\bB\cdot\bmm$
\end{widetext}
\bibliography{references}
\end{document}